\documentclass[aps,preprintnumbers,showpacs,showkeys,nofootinbib,a4paper]{revtex4}
\usepackage[utf8]{inputenc}
\usepackage{t1enc}
\usepackage{subfigure}
\usepackage{graphicx}
\usepackage{dcolumn}
\usepackage{epsfig}
\usepackage{bm}
\newcommand{\eq}{\begin{eqnarray}}
\newcommand{\en}{\end{eqnarray}}
\newcommand{\subfloat}{\subfigure}

\newcommand{\ba}[1]{\begin{eqnarray} \label{(#1)}}
\newcommand{\ea}{\end{eqnarray}}

\newcommand{\newc}{\newcommand}
\newc{\lra}{\leftrightarrow}
\newc{\beq}{\begin{equation}}
\newc{\eeq}{\end{equation}}
\newc{\barr}{\begin{eqnarray}}
\newc{\earr}{\end{eqnarray}}
  
    \def\sbf{\mbox{\boldmath $\sigma$}}

\begin{document}

\topmargin -0.50in
\title {Calculated event rates for Axion Detection via Atomic and Nuclear Processes}

\author{John D. Vergados}
\affiliation{ University of Ioannina,  Ioannina, Gr 451 10, Greece
and Center for Axion  and  Precision  Physics Research, IBS,
Daejeon 34051, Republic of Korea}

\author{Paraskevi C. Divari}
\affiliation{Department of Physical Sciences and Applications,
Hellenic Military Academy, Vari 16673, Attica, Greece}

\author{Hiroyasu Ejiri}
\affiliation{Research Center for Nuclear Physics, Osaka
University, Osaka 567-0047, Japan}

\begin{abstract}
The possibility of detection of 5.5 MeV and 14.4 keV solar axions
 by observing axion induced nuclear and atomic  transitions is investigated.
  The presence of nuclear  transitions  between spin orbit partners can
  be manifested by the subsequent de-excitation via gamma ray emissions.
  The transition rates  can also be studied in the context of radiative axion
  absorption by a nucleus. The elementary interaction is obtained in  the context
  of the   axion-quark  couplings  predicted by existing axion models.
   Then these couplings  will be transformed to the nucleon level utilizing
   reasonable existing   models, which lead  to  effective transition operators.
    Using these operators we calculate  the needed nuclear matrix elements employing
     wave functions obtained  in the context of the nuclear shell model. With these
     ingredients we discuss possibilities of experimental observation of
      the axion-induced nuclear gamma-rays.  In the second part we will
      examine the axion induced production of X-rays (axion-photon conversion)
       or ionization from deeply bound electron orbits. In this case the axion
       electron coupling is predicted by existing axion models, no renormalization is needed.
        The experimental signal is the observation of directly produced electrons and/or the emission
         of hard X-rays and Auger electrons, following the de-excitation of the final atom.
         Critical discussion is made on the  experimental feasibility of detecting the solar
          axions by using multi-ton scale NaI detectors.
\end{abstract}


\date{\today}

\maketitle

\section{Introduction}
In the standard model there is a source of CP violation from the phase in the Kobayashi-Maskawa mixing
matrix. This, however, is not large enough to explain the baryon asymmetry observed in nature.
Another source is the phase in the interaction between gluons ($\theta$-parameter), naively  expected to be of order unity.
The non observation, however, of elementary electron dipole moment, set limits its value to be $\theta\le10^{-9}$.
This has been known as the strong CP problem.
A solution to this problem has been the P-Q (Peccei-Quinn) mechanism.
In extensions of the Standard Model (S-M), e.g. two Higgs doublets, the Lagrangian has a global
 P-Q chiral symmetry $U_{PQ}(1)$, which is spontaneously broken, generating a Goldstone boson, the
 axion (a). In fact
the axion has been proposed a long time ago as a solution to the strong CP problem \cite{PecQui77}
resulting to a pseudo Goldstone Boson \cite{SWeinberg78,Wilczek78}. The two most widely cited models
of invisible axions are the KSVZ (Kim, Shifman, Vainshtein and Zakharov) or hadronic axion models
 \cite{KSVZKim79,KSVZShif80}
and the  DFSZ (Dine,Fischler,Srednicki and Zhitnitskij) or GUT
axion model \cite{DineFisc81,DFSZhit80}.
 This also led  to the interesting scenario of the  axion being a candidate for dark matter in the
 universe \cite{AbSik83,DineFisc83,PWW83} and it can be searched for by real
experiments \cite{ExpSetUp11b,Duffy05,ADMX10,IrasGarcia12}. For a recent review see \cite{Bibber16}.
The relevant phenomenology has  also been reviewed \cite{PhysRep16}.

Axions could be generated  mainly  by Bremsstrahlung processes and
Compton scattering \cite{Kaplan85}, originally proposed for star
cooling via axion emission, see e.g.  \cite{AxionBrem97} and a
more recent work  \cite{AxionBrem19} for Bremsstrahlung processes
and   Compton scattering studied a long time ago
\cite{AxionCompton97} and more recently \cite{AxionBremCom11}.
Also generated via the Primakoff conversion of  photons to axions
in the presence of a magnetic field \cite{Primakoff51}. All these
yield a  continuous axion spectrum, the Primakoff conversion with
an average energy around  4 keV \cite{CAST05}. They can also be
generated  by nuclear transitions, which involve monochromatic
axions, e.g.  14.4 keV in the case  of the
 de-excitation of $^{57}$Fe* and 5.5 MeV in the case $p+d\rightarrow ^3_2$\mbox{He*}.
 Axions produced in nuclear processes are mono-energetic because their
energies correspond  to the transition energies to specific
nuclear states.  These solar axions can be emitted and escape from
the solar core due to the very weak interaction between the axion
and matter.

In the   Intensity Frontier of physics a search for new light particles,
including axions among others, is growing.  Generally speaking axions can be
 produced in an environment of intense photon production. The axion sources
 considered so far are cosmological or in the sun.
 Such environment  on earth can be  the core of a nuclear
    reactor, see e.g. the recent works \cite{Papoulias21,deNLeeLee21}.
    It is, of course, well known that neutrinos were first detected in reactors
    and the most accurate neutrino oscillation parameters were determined using reactor neutrinos.
    It is amusing to think that history may repeat itself, i.e.  reactors may  have  the potential
    to produce axion like particles (ALPs) of sufficient intensity through
    Primakoff conversion, Compton-like processes, nuclear transitions  as well as dark photons. The (ALPs)
    can subsequently interact with the material of a nearby detector via the standard processes,
     i.e. the inverse   Primakoff effect, the inverse Compton-like scatterings and  via axio-electric absorption.
     It is claimed \cite{Papoulias21,deNLeeLee21}, that  reactor-
    based neutrino experiments have a high potential to test ALP-photon couplings
    and masses.

 Searches for solar axions have been carried out with
various experimental techniques: magnetic helioscopes
\cite{CAST09,CAST14}, low temperature bolometers~\cite{Bolom13}
and thin foil nuclear targets~\cite{ThinFoil07}.
 CUORE(Cryogenic Underground
Observatory for Rare Events) \cite{CUORE04,CUORE05,CUORE17,Alduino17,CUOREColl13} as well as
 the Majorana Demonstrator  \cite{Majorana16} which are designed to search for neutrinoless double beta
decay ($0\nu\beta \beta$) using  very low background  detectors
are used for the axion searches. The Ge detectors in  GERDA
\cite{GERDA13} can also be used to search for dark matter weakly
interacting massive particles (WIMPs) and  axion searches
\cite{LiAvigWang16,Avignone09,CoGeNT11}.

As we have mentioned high energy axions can be produced in the
sun, e.g. in the keV range, by the de-excitation of $^{57}$Fe*
\cite{SolFlux20} or even at the MeV range
\cite{AxionBorex12,CAST10,BHHousLi20}   via the reaction : \beq
p+d\rightarrow ^3_2\mbox{He}+a(5.5)\mbox{MeV together with the
observed }p+d\rightarrow ^3_2\mbox{He}+\gamma(5.5)\mbox{MeV}
\label{Eq.SolarFlux} \eeq Both solar axions mentioned above can be
exploited for axion detection. Those in the energy range of Eq.
(\ref{Eq.SolarFlux}) are suitable for the study of axion induced
nuclear transitions in the laboratory, while the keV axions for
atomic processes.

To this end  in section II we will discuss the particle model
needed in understanding these processes. In section III we discuss
how one obtains a set of axion-nucleon couplings   $g_{aN}$, going
from the quark to the nucleon level, needed in constructing the
effective nuclear transition operators. In section IV we will
briefly discus the experimental issues involved in the detection
of axions in large detectors involving materials like NaI. In
section V we will derive the cross sections needed for the rates:
A) in the case of spin induced nuclear  transitions, B) the
radiative axion absorption by  nuclei and C) axion induced weak
process induced by axion absorption. Then in section VI  we will
briefly discuss the nuclear model needed to evaluate the cross
sections for cases A) and B) above for the Na and I nuclei. In
section VII we  will present the obtained nuclear matrix elements
for the nuclei of interest and we will exhibit  our results for
the relevant cross sections and the  obtained event rates.\\
After that we will utilize the Fe$^*$-57 14.4 keV source, which is
ideal to study  axion induced atomic processes. Thus, after
determining the axion-electron coupling from the Borexiono
experiment in section VIII, we will calculate in sections IX and X
the cross sections for axion induced atomic transitions and
ionization from deeply bound orbits. In this case one expects good
experimental signatures, in addition of observing the primary
electrons one expects signals of hard X-rays and Auger electrons,
following the atomic de-excitation.

\section{The particle model}
\label{sec:pmodel}

We remind the reader that the axion, $a$, is a pseudoscalar particle with a coupling to fermions.\\
i) The axion  coupling to the electron can be described by a Lagrangian of the form:
\beq
{\cal L}=  \frac{g_e}{f_a}i\partial_{\mu}a\bar{\psi}_e({\bf p}',s)\gamma^{\mu}\gamma_5\psi_e({\bf p}, s)
\eeq
where $g_e$ is a coupling constant and $f_a$ a scale parameter with the dimension of energy. $ g_e$ can be
 calculated in various axion models. For an axion with mass $m_{a}$ it easy to show that  in the non relativistic limit:
\begin{itemize}
    \item the time component $\mu=0$ is given by:
    \beq
    {\cal L}=\langle \phi|\Omega|\phi \rangle,\quad\Omega=\frac{g_e m_a}{2 f_a}\frac{\sbf.{\bf q}}{m_e},\quad{\bf q=p'-p}
    \label{Eq:AxionelInt}
    \eeq
    which is negligible for $m_a<< m_e$.
    \item The space component, $\mu\ne 0$,
    \beq
    {\cal L}_{aee}=\langle \phi|\Omega|\phi \rangle,\quad\Omega=\frac{g_e}{2 f_a}\sbf.{\bf q},\quad{\bf q=p'-p}
    \label{Eq:spinae}
    \eeq
    where ${\bf p}$ and ${\bf p}'$ are the initial and final electron momenta,  $f_a$  the axion
    decay constant and $\sbf$ the spin of the electron.\\
\end{itemize}

 ii) The axion  coupling to the quarks is similarly given by:
\beq {\cal L}=  \frac{g_q}{f_a}i\partial_{\mu}a\bar{\psi}_q({\bf
p}',s)\gamma^{\mu}\gamma_5\psi_q({\bf p}, s), \label{Eq:FundInt}
\eeq where $g_q$ is the coupling constant. The space component,
$\mu\ne 0$, in the non relativistic limit is given by \beq {\cal
L}=\langle \phi|\Omega|\phi \rangle,\quad\Omega=\frac{g_{aq}}{2
f_a}\sbf.{\bf q},
\label{Omegaq}
\eeq
with $\sbf$ the Pauli matrices and  $\phi$  the quark wave function and ${\bf q}$ is the axion momentum.

We will concentrate on the last term involving the operator
$g_{aq}\sigma$. The quantities $g_{aq}$ can be evaluated at
various axion models. \beq \mbox{me}_q=\langle q|\left (
g_{aq}^0+g_{as}+g_{aq}^3\tau_{3}\right ) \sigma |q\rangle, \eeq
where we have ignored the contribution of heavier quarks and,
thus, the   isovector and isoscalar components become \beq
g_{aq}^3=\frac{1}{2}(g_{au}-g_{ad}),\quad
g_{aq}^0=\frac{1}{2}(g_{au}+g_{ad}). \eeq Some authors use the
notation $c_q$ instead of $g_{aq}$.

The scale parameter $f_a$ has been  related to the axion mass
$m_a$ via the relation: \beq \frac{1}{f_a}
=\frac{m_a}{\Lambda^2_{QCD}}, \label{Eq:MafaLambdaQCD} \eeq with $
\Lambda_{QCD}=218\mbox{ MeV}$. For the determination of
$\Lambda_{QCD}$ see Particle Data Group \cite{LambdaQCD16}. More
recently, however, a model independent relation has been provided
\cite{CorGhVill18} \beq m_af_a=5.691(51)\times 10^{3}\mbox{MeV}^2
\eeq We will employ the expression \beq m_af_a\approx 6000
\mbox{MeV}^2 \label{Eq:mafaNew} \eeq Taking $f_a$ to be
$10^{10}$MeV then  $m_a=0.6$eV.

\section{The effective axion nucleon couplings}

In this section we follow the procedure discussed in a previous
work  \cite{SolFlux20} and references therein. For  the benefit of
the reader we are going to give a summary here.

 In going from the
quark to the nucleon level one can follow  a procedure analogous
for the determination of the nucleon spin from that of the quarks
\cite{QCDSF12,LiThomas15}. The
 matrix element at the nucleon  level can  be written as:
\beq \mbox{me}_N=\langle N |\left ( g_{aq}^0(\delta_0-\Delta
s)+g_{as}\Delta s+g_{aq}^3\tau_{3}\delta_1\right )\sigma |
N\rangle, \eeq where \beq \delta_0=(\Delta u+\Delta d+\Delta
s),\,\delta_1=(\Delta u-\Delta d), \label{Eq:a0a1L} \eeq
$\delta_0-\Delta s=\Delta u+\Delta d$. The quantities $ \Delta
u,\, \Delta d$ and $ \Delta s$ have been obtain previously
\cite{SolFlux20}. Alternatively these quantities can be expressed
in terms of the quantities $D$ and $F$  defined by Ellis
\cite{JELLIS}
$$\delta_1=F+D,\,\delta_0=3F-D+3\Delta s.$$
The quantity $\delta_1$ is essentially fixed by the axial current to be approximately 1.27
 of the vector current. No such constraint exists for the isoscalar part, so for that we have
  to rely on models. For the quantities $D$ and $F$ one can use experimental information \cite{JELLIS}.
   Thus, e.g., from hyperon beta decays and flavor SU(3) symmetry one gets
$$\frac{3F - D}{\sqrt{3}}= 0.34 \pm 0.02.$$
Different  approach has been suggested by Kaplan \cite{Kaplan85},
based on the DFSZ \cite{DineFisc83, DFSZhit80} axion, which is
analyzed in some detail in  ref. \cite{SolFlux20}.

 On the other hand measurements of
$\nu p$  and $\nu \bar{p}$ elastic scattering   of the recent MicroBooNE experiment \cite{MicroBooNE17}
 indicate that $\Delta s = \pm 0.036\pm 0.003$. This is also discussed in \cite{SolFlux20}.
It is, however, concluded \cite{SolFlux20} that the relevant couplings cannot be extracted
from experiment in the presence of both proton and neutron components in the nuclear wave functions.

So we have to rely on the above models. The obtained results are found in Tables 2 and 3 in
 ref. \cite{SolFlux20}. From these the favorite choice is
\barr
C_p&=&0.0663,C_n=0.0663\mbox{ for }\tan{\beta}=1 \mbox{ MODEL E},\nonumber\\C_p&=&0.2712,C_n=-0.1248 \mbox{ for }\tan{\beta}=10 \mbox{ MODEL F}.
\label{Eq:bestF}
\earr
where $C_p $ and  $C_n$ are hadronic couplings for protons and neutrons
 respectively and $\tan{\beta}$ is the ratio of the vacuum expectation
 values of the two Higgs doublets (see  ref. \cite{SolFlux20}).
 Theoretical  preference is given to the higher value of $\tan{\beta}$.
 It also yields an isovector coupling  larger than the isoscalar, which is
 sometimes useful in obtaining the ratio of axion to photon production
 rates (the photon production being  dominated by the isovector transition).

The effective axion-nucleon coupling can  be cast in the form:
\beq H_{aNN}=\frac{1}{2f_a}\left ( g_{aN}^0+g_{aN}^3\tau_{3}\right
)\sbf\cdot {\bf q},\,g_{aN}^0=\frac{1}{2}\left (C_p+C_n \right
),\,g_{aN}^3=\frac{1}{2}\left (C_p-C_n \right )
\label{Eq:effANucCoup} \eeq

Before proceeding further with the evaluation of the
nuclear and atomic matrix elements of interest we would
like to consider the experimental situation in detecting the
 resulting signals using a large NaI  detector, already developed and in use for rare event detection.

\section{Experimental aspects of the axion search by NaI}

We will now briefly present some remarks on experimental aspects
of the axion searches by nuclear interactions. The 5.5 MeV solar
axion has been well investigated by using the Borexino detector
with 278 tons of $\mbox{C}_9 \mbox{H}_{12}$ scintillation
detector. Among the interactions studied are the
axion-electron/photon interaction. Then, the upper limits for the
product of the axion flux and the cross section is around $4.5
\times 10 ^{-3}$SNU, much smaller, by almost   3 orders of
magnitude, than the rates for $^8$B neutrinos without oscillation.
Note the axion energy and the $^8$B neutrino energy are similar.
The limits on the axion couplings are derived as $|g^3_{aN} \times
g_{aE} | \leq 5.5 \times  10^{-13} $. It is noted that the
coupling $|g^3_{aN}|^2$ is studied from the axion nuclear
interaction. Experimental signals from the axion-induced
photo-production are ultra rare 5.5 MeV $\gamma$-rays. The signal
rate $N _S $ per t y, with t being the target nuclear mass in
units of tonne and y being year, is given as \beq N _S = 600 N
A^{-1} \label{Eq:NS} \eeq where $N$ is the signal rate per
$10^{27}$  nuclei  and $A$ is the mass number. So we get $N_S
=0.5$ and $N_S =0.005$ for $N =0.1$ and 0.001 in the case of
$A=127$ for Iodine  nuclei. We note the signal rates are just of
the same order of magnitude as the rates of the neutrinoless
double beta decays (DBDs), respectively, for the inverted-mass and
normal-mass hierarchies (IH and NH ) for typical DBD nuclei. The
signal energy of 5.5. MeV is similar to that of the  DBD, i.e. of
around 3 MeV. Then one may have to use  multi-ton or even
multi-hundred ton scale detectors as used for the IH and NH
neutrino-mass DBD experiments. It is crucial for the rare-event
search to reduce (BG) background counts $B$ per t y to the level
of $B \leq N_S$ . Backgrounds due to radio-active impurities to be
considered are the $\beta-\gamma$ rays from $^{214}$ Bi and
$^{208}$ Tl. They are known to be serious in case of DBD
experiments with the signals around or below 3 MeV, but are not
serious in the present case of the signals above 5 MeV. Alpha
particle backgrounds are separated from the $\gamma$ -ray signals
by a pulse-shape discrimination. The background event rate for the
$^8$B solar-neutrinos $B_{\nu}$ /t y is expressed
\cite{Ejiri05,EjSuhZub19,EjZub17,EjEll14,EjEll19} as \beq B_{\nu}=
0.15 \times E \Delta, \eeq where $E$ is the energy in units of
MeV, i.e. 5.5 in the present case, and $\Delta$ is the ratio of
the energy window to E. The 5.5 MeV signal rate for the realistic
axion is too small to be detected beyond the background neutrinos.
Thus the search for such solar axion is not realistic at present.

In case of the 14.4 keV axion search, the BG rate in the energy
region is  of the order of or larger than 1/(t y) because of all
kinds of radio-active impurities in addition to the solar-neutrino
BG. The background level of  current low-background detectors
being used for dark-matter searches is of the order of $10^5/(t
y)$. The signal to be considered is the 14.4 keV one from the
axion-induced atomic electron.  Then on may search for such axions
in the 5eV region, which is interesting but not yet well
investigated, by improving the background level by a few orders of
magnitude. Thus it is interesting to discuss the rate as given in
sections IX and X.

\section{Cross sections for nuclear process}

Using the effective axion nucleon coupling given by Eq,
(\ref{Eq:effANucCoup})
 the nuclear matrix element (ME) is of the form:
\beq
\mbox{ME}=\langle \Psi_f|{\bf O}\cdot {\bf q}|
\Psi_i\rangle,\,{\bf O}=\sum_k\left ( g_{aN}^0+g_{aN}^3\tau_{3}(k)
\right )\sbf(k) \label{Eq:GPhoto} \eeq with the summation
involving all nucleons. In the case of the MODEL F of Eq.
(\ref{Eq:bestF}) we find $g_{aN}^0 =0.073$ and $g_{aN}^3=0.197$.
This can trivially be written in the proton-neutron if dictated by
nuclear physics.

We will consider two processes\\
i) nuclear excitation: $$ a+A(N,Z)\rightarrow A(N,Z)^*$$
followed by de-excitation emitting  $\gamma$ rays and \\
ii) axion photoproduction:
$$ a+A(N,Z)\rightarrow A(N,Z)^*+\gamma$$

Note that  the signals of i) and ii) are  5.5~MeV gammas, which
are not separated. Thus one may consider one of them, the one with
the larger signal rate.

\subsection{Axion induced nuclear excitation}

The capture cross section associated with  Fig.  \ref{fig:Nga}(a)
takes the form \beq \sigma(E_{a})=\frac{1}{\upsilon}\frac{1}{2
E_a} 2 \pi \delta(E_a-\Delta) {\cal M}^2 \eeq where $\Delta$ is
the  energy  transferred to the nucleus. If the nuclear recoil
energy is ignored $\Delta$ corresponds to the nuclear excitation
energy. ${\cal M}$ is the invariant amplitude, essentially the
nuclear matrix element, and $1/E_a$ is the usual normalization of
a scalar field. Summing over the final substates $M_f$ and
averaging over the initial substates $M_i$ we find: \beq
\sigma({E_{a}})=\frac{E_a}{q c^2}\frac{1}{2E_a} 2 \pi
\delta(E_a-\Delta) \frac{1}{(2f_a)^2}\frac{1}{3}q^2\frac{1}{2
J_i+1}\langle J_f||{\bf O}||J_i\rangle^2 \eeq where $\langle
J_f||{\bf O}||J_i\rangle $ is the spin-reduced matrix element of
the operator ${\bf O} $ discussed above.  If the spectrum of the
source is continuous given by a distribution $\rho(E_a)$ the
total cross section is given by:
 \beq
  \sigma(\Delta)=\int
dE_a \rho(E_a)\sigma(E_{a})=  \frac{\pi}{12f^2_a}\rho(\Delta)
\sqrt{\Delta^2-m^2_a} \frac{1}{2 J_i+1}\langle J_f||{\bf
O}||J_i\rangle^2 \eeq with $m_a$ the axion mass.

In the present case, however, the spectrum resulting from the
process given by Eq. (\ref{Eq.SolarFlux}) is monochromatic and we
have a resonance. So to proceed further we assume a Breit-Wigner
shape : \beq \rho(E)=\frac{1}{2
\pi}\frac{\Gamma}{(E-E_a)^2+\Gamma^2} \label{Eq:Rhog} \eeq with a
small width $\Gamma$ determined by the production rate centered at
the monochromatic energy, which is   tiny. In fact  in the case of
14.4 keV solar axion  for $m_a=5$ eV one finds \cite{SolFlux20}
that the width $\Gamma=10^{-24}$ eV. The situation for 5.5 MeV
$^3$He solar axions, Eq. \ref{Eq.SolarFlux}, is similar. So it is
reasonable to assume a flat distribution centered around $E_a$,
that is: \beq \rho(E_a)=\left \{ \begin{array}{cc}\frac{1}{ 2
\kappa \Delta},& (1-\kappa)\Delta \le E_a \le (1+\kappa) \Delta,\,
0<\kappa< 1\\0&\mbox{ otherwise}\\ \end{array} \right .
\label{Eq:Rhosq} \eeq The distribution of Eq. (\ref{Eq:Rhosq}) is
normalized the same way as that of Eq. (\ref{Eq:Rhog}), that is to
unity.  We thus get
 \beq
 \sigma(\Delta)= \sigma_{aN}  \frac{1}{2 J_i+1}\langle J_f||{\bf O}||J_i\rangle^2,\,\sigma_{aN}=\frac{\pi}{12 f_a^2} \frac{\sqrt{\Delta^2-m^2_a}}{2 \kappa \Delta}.
 \label{Eq:DeltaE}
 \eeq
 For axions with a mass much smaller than its energy we find according to (\ref{Eq:mafaNew})
 \beq
  \sigma_{aN}=  \frac{1}{2 \kappa}\times 1.00\times 10^{-42}\mbox{cm}^2({m_a}/{0.6 \mbox{eV}})^2
 \label{Eq:DeltaE2}
 \eeq
 or
 \beq
  \sigma_{aN}\approx  \frac{1}{2 \kappa}\times 6.28\times \times 10^{-41}\mbox{cm}^2({m_a}/{5 \mbox{eV}})^2
 \label{Eq:DeltaE3}
 \eeq

\subsection{Axion photoproduction}

This process is reminiscent of the old pion photoproduction,
see e.g. \cite{EricRho72} for a review and \cite{JDV75} for nuclear
applications. It involves the absorption of negatively charged   pion absorbed by a nucleus
from a bound state in  an atomic like orbit. Since the pion is charged the final nucleus
has a different charge than the original one.  There exist, of course,
differences in relation to the axion having to do with its coupling,
the energy and the fact that  the axion  is electrically neutral, so that in this case
the final nucleus coincides with the original one. Thus   one expects the
 prospect  an elastic  coherent process as well, with a very large cross section.

\begin{figure}[htbp]
 \begin{center}
   \subfloat[]
 {
    \includegraphics[width=0.2\textwidth]{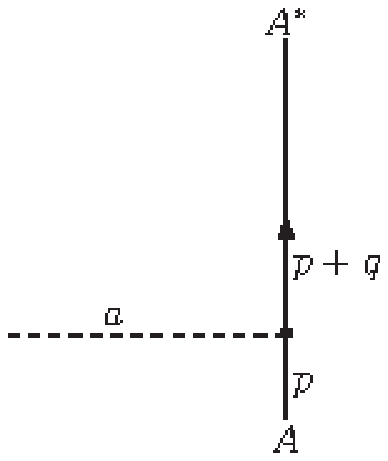}
}
 \subfloat[]
 {
    \includegraphics[width=0.4\textwidth]{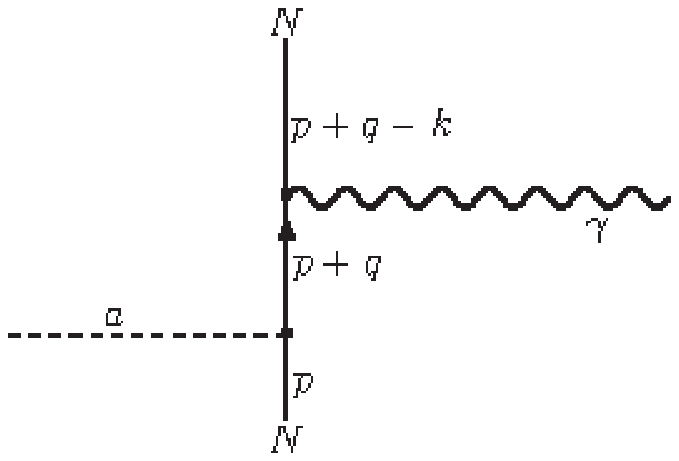}
 }
\\
  \caption{(a) An axion is absorbed by a nucleus leading to an excited nuclear state.
  (b ) A Feynman diagram relevant for axion photoproduction. In both cases
   the axion vertex we encounter  the coupling given by Eq, (\ref{Eq:effANucCoup}),
    while the EM vertex in (b) we have the usual interaction involving the
     nuclear magnetic moment. p,  q and k are the initial nucleon, axion and photon 4-momenta respectively. In both cases of (a) and (b),
     the out-put signal energies are the same 5.5 MeV. }
    \label{fig:Nga}
    \end{center}
\end{figure}

The relevant elementary amplitude at the nucleon level is given in Fig.   \ref{fig:Nga}(b).

 The non relativistic reduction of this amplitude leads to an effective operator with the structure
\beq
\frac{\mu_B(N)}{f_a 2E_a}\left (\Omega_A+\Omega_B+\Omega_C \right )
\eeq
with $\mu_B(N)$ the Bohr magneton for the nucleon and
\barr
\Omega_A&=&\left (G^0_{aN}+G^3_{aN}\tau_3\right )({\bf q}\times i{\bf k})\cdot \epsilon_{\lambda},\nonumber\\
\Omega_B&=&\left (G^0_{aN}+G^3_{aN}\tau_3 \right )\sbf \cdot(i{\bf
k} \times \epsilon_{\lambda})E_a,\,\Omega_C=\left
(G^0_{aN}+G^3_{aN}\tau_3 \right )\sbf \cdot\left ((
\epsilon_{\lambda} \times {\bf k}) \times{\bf q} \right )
\label{Eq:PhoprodOp} \earr where $ \epsilon_{\lambda}$ and ${\bf
k}$ the polarization and momentum of the photon, $E_a$ and {\bf q}
the energy and momentum of the axion. Furthermore \beq
G^0_{aN}=\frac{1}{2}(g_s(p) C_p+g_s(n)
C_n),\,G^3_{aN}=\frac{1}{2}(g_s(p) C_p-g_s(n) C_n). \eeq Using the
values of Eq. (\ref{Eq:bestF}), MODEL F, we obtain: \beq
G^0_{aN}=0.781,\,G^3_{aN}=0.734 \eeq We notice that in this case
we could have elastic transitions involving the spin independent
operator $\Omega_A$. So in this case, proceeding as above, we can
write \barr
{\cal M}^2&=&\frac{\mu^2_B(N)}{f^2_a 4E_a^2} k^2E_a^2
\left(\mbox{ME}_{el}^2+\mbox{ME}^2_{i\rightarrow f}\right )
\mbox{ with }\nonumber\\\mbox{ME}_{el}^2&=&\left \{ \left(G^0_{aN}A+ G^3_{aN}(Z-N)\right )^2\frac{q^2}{E_a^2}
(1-(\hat{k}\cdot \hat{q})^2)+\frac{1}{3}\frac{1}{2 J_i+1}\langle J_i||(G^0_{aN}+G^3_{aN}\tau_3 )
\sbf||J_i\rangle^2 \left(1+\frac{q^2}{E_a^2} \right ) \right \},\nonumber\\
\label{Eq:elasticNucleus} \earr where $k$ is the photon energy,
which in this case is equal to the axion energy. {\bf q} is the
axion momentum and $q^2=E^2_a-m^2_a\approx E^2_{a}$ since the
axion mass is very small. The first term in the last expression is
reminiscent of the coherent contribution of all nucleons in the
more familiar case of the spin independent contribution in the
WIMP-nucleus scattering (WIMP stands for weakly interacting
massive particle).  In deriving this expression we have summed
over all polarizations and final magnetic substates and sum over
the initial m-sub-states. The last sum includes the contribution
of both  $ \Omega_B$ and $ \Omega_C$, which contribute only if
$J_i\ne 0$. When averaged  over the photon momentum directions the
contribution of the first term is reduced by 2/3.

For transitions to excited states we simply write \beq
\mbox{ME}_{i\rightarrow f}^2=\left \{ \frac{1}{3} \frac{1}{2
J_i+1}\langle J_f||(G^0_{aN}+G^3_{aN}\tau_3
)\sbf||J_i\rangle^2\left(1+\frac{q^2}{E_a^2} \right ) \right \}
\eeq The differential cross section, neglecting the energy of the
recoiling nucleus, becomes \beq
d\sigma(E_{a})=\frac{1}{\upsilon}\frac{1}{2 k} \frac{1}{2 E_a}
2\pi \delta(E_a-\Delta-k)\frac{1}{(2 \pi)^3}d^3k {\cal M}^2,\,
\Delta\mbox{ being  the excitation energy}\Rightarrow \eeq \barr
d\sigma(E_{a})&=&\frac{E_a}{q c^2}\frac{1}{2 k} \frac{1}{2 E_a} \frac{1}{(2 \pi)^3}\frac{\mu^2_B(N)}{f^2_a 4E_a^2} k^2E_a^2\left (\mbox{ME}_{el}^2+\sum_f \mbox{ME}_{i\rightarrow f}^2\right )  2\pi \delta(E_a-\Delta-k)4 \pi k^2 dk \nonumber\\
&=&\frac{E_a}{q c^2}\frac{1}{4 \pi}\frac{\mu^2_B(N)}{f^2_a 4E_a^2} \left (\mbox{ME}_{el}^2+\sum_f \mbox{ME}_{i\rightarrow f}^2\right )k E_a \delta(E_a-\Delta-k)k^2 dk\Rightarrow
 \earr
 \beq
 \sigma(E_{a})=\frac{E_a}{q c^2}\frac{1}{4 \pi}\frac{\mu^2_B(N)}{f^2_a 4E_a^2} \left (\mbox{ME}_{el}^2+\sum_f \mbox{ME}_{i\rightarrow f}^2\right )E_a(E_a-\Delta)^3,
 \eeq
 $ E_a-\Delta$ is, of course, equal to the photon energy.
 The cross section  can also  be expressed  in terms of the axion mass:
  \beq
 \sigma(E_{a})=\sigma_0(a\gamma)\left (\mbox{ME}_{el}^2+\sum_f \mbox{ME}_{i\rightarrow f}^2\right )\left( 1-\frac{\Delta}{E_a}\right )^3,\quad\sigma_0(a\gamma)= \frac{E_a}{q c^2}\frac{1}{4 \pi}\frac{E_a^2\mu^2_B(N)}{ 4}  \frac{1 }{f_a^2}\hspace{3pt}
 \eeq
 Thus the scale of the cross section for axion photoproduction for $f_a=10^7$ GeV, which corresponds to $m_a=0.6$ eV, is
 \beq
  \sigma_{a\gamma}=6\times 10^{-50}\mbox{cm}^2\left (E_a/5.5 \mbox{MeV}\right )^2\left (m_a/0.6 \mbox{eV}\right )^2
   \label{Eq:sgma0RadCap}
 \eeq
 or
 \beq
  \sigma_{a\gamma}\approx 4 \times 10^{-48}\mbox{cm}^2\left (E_a/5.5 \mbox{MeV}\right )^2\left (m_a/5 \mbox{eV}\right )^2
   \label{Eq:sgma0RadCap2}
 \eeq
 This seems suppressed compared, e.g., to the axion-nucleus cross section discussed above. This is not surprising,
 since the axion energy of 5.5 MeV is much smaller then the nucleon mass
 $E_a^2\mu^2_B(N)<<1$.
 This can be overcome by the fact that  many final states below $E_a$  can contribute.
Furthermore the elastic term contains a large coherent
contribution, which is proportional to $A^2$. We should also keep
in mind that in this case  the cross section includes the
radiative part as well.

\subsection{Weak process induced by axion  absorption}

\begin{figure}[htbp]
    \center
    \includegraphics[width=0.7\textwidth]{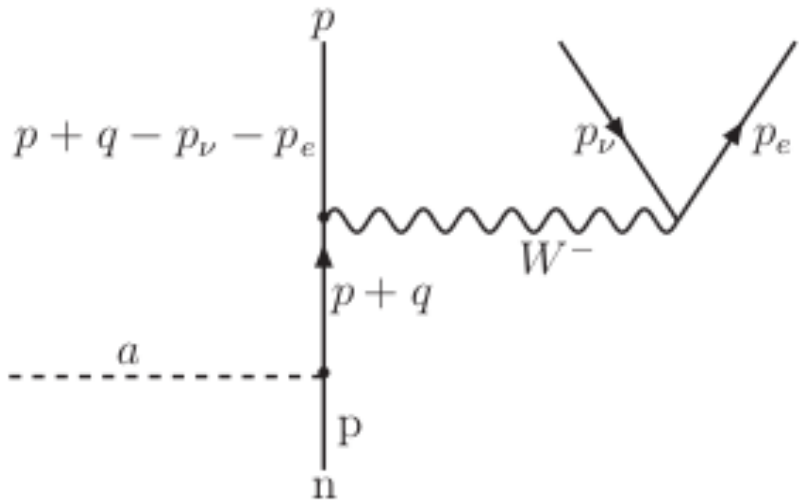}
    \caption{A Feynman diagram relevant for axion induced beta decay}
    \label{fig:NgaBeta}
\end{figure}

This can occur in a process given by  the Feynman diagram shown in Fig. \ref{fig:NgaBeta}.
Proceeding as in the previous subsection we find the following possibilities;
$$\Omega=\frac{G_F}{\sqrt{2}}\frac{1}{f_a 2 E_a} C_nE_a\left (((N\tau_+N) j_0 +(N\tau_+\sbf N)\cdot {\bf j})  -(N\tau_+ N) {\bf j}\cdot \frac{{\bf q}}{E_a}
-( N \tau_+ \sbf N) \cdot \frac{(-i{ \bf q}\times {\bf
j})}{E_a}\right ) $$ where $C_n$ is the appropriate axion-neutron
coupling appropriate for negaton decay.
$$j_{\mu}=\bar{e}\gamma_{\mu}(1-\gamma_5)\nu_e$$
is the usual weak leptonic current. The first two terms are
exactly the Fermi and the GT matrix elements   encountered in the
allowed beta decay. The other two contribute similarly,  since,
for a light axion, $q^2/E^2_a\approx 1$. So one encounters the
contribution $2 \left (\mbox{ME}^2_F+\mbox{ME}^2_{GT} \right )$.

The differential cross section takes the form \beq d
\sigma_{a\beta}=\frac{1}{\upsilon}\frac{1}{2 E_a}\left (
\frac{G_F}{\sqrt{2}}\frac{1}{f_a  } C_n \right )^2 2 \left
(\mbox{ME}^2_F+\mbox{ME}^2_{GT} \right ) m^2(p_e,p_{\nu}) 2 \pi
\delta(E_a-\Delta-E_{e}-E_{\nu})\frac{d^{3}{\bf p}_e}{(2
\pi)^3}\frac{d^{3}{\bf p}_{\nu}}{(2 \pi)^3} \eeq where $\Delta$ is
the nuclear energy difference involved, $m^2(p_e,p_{\nu}) $ the
standard leptonic matrix element. The total cross section can be
cast in the form: \beq \sigma_{a\beta}=\sigma_{0a\beta} \left
(\mbox{ME}^2_F+\mbox{ME}^2_{GT} \right )\frac{1}{m_e^5}\int p_e^2
dp_e p^2_{\nu}d p_{\nu} m^2(p_e,p_{\nu}) 2 \pi
\delta(E_a-\Delta-E_{e}-E_{\nu}) \label{Eq:betacros} \eeq with
$$ \sigma_{0a\beta}=\frac{1}{2}\frac{1}{(2 \pi)^3}\left ( G_F m_e^2\right )^2\frac{m_e}{E_a}\frac{C^2_n}{f^2_a}$$
Thus for $f_a=10^{7}$ GeV we find:
$$ \sigma_{0a\beta}=1.1 \times 10^{-64}\mbox{cm}^2(E_a/5.5 \mbox{MeV})^{-1}$$
Since this corresponds to $m_a=0.6$ eV we find
$$ \sigma_{0a\beta}=1.1 \times 10^{-64}\mbox{cm}^2(E_a/5.5 \mbox{MeV})^{-1}(m_a/0.6 \mbox{eV})^2$$
or
$$
\sigma_{0a\beta}=7.6 \times 10^{-63}\mbox{cm}^2(E_a/5.5 \mbox{MeV})^{-1}\left (\frac{m_a}{5 \mbox{eV}}\right )^2
$$
This is much smaller than the $ \sigma_{0\gamma}$ discussed above, which is not
surprising due to the fact that we now deal with a weak process. The scale of
 this cross section has  perhaps been  underestimated, scaling it with $m_e$ rather
 than the energy available in the decay. This, however, is not lost, since it will
 show up in the integral involved in Eq. (\ref{Eq:betacros}).

In any case it will be of interest to compare this process with the  standard
 beta decay. Leaving aside the nuclear matrix elements we find:
\beq \frac{R_{0a,\beta}}{R_{0\beta}}=\frac{N_A
\sigma_{0a\beta}\Phi(E_a)}{\frac{1}{2
\pi^3}(G_Fm_e^2)^2m_e}\approx \frac{10^{27}\times7.6\times
10^{-63}\times 1.3 \times 10^{9}\mbox{s}^{-1}}{0.016\times 9.62
\times 10^{-24}\times 7.8 \times 10^{20}}=8.3 \times 10^{-23} \eeq
Experimentally, the 5.5 MeV state, if excited , decays immediately
by gamma emission with $10^{-15}$~sec, and thus the weak decays
are not considered experimentally as in many other cases.

Let us hope that a target with $N_A>>10^{27}$ is possible\footnote{In high energy experiments searching, e.g., for proton decay, about $10^{32}-10^{33}$ nuclei have been accumulated, but in this case the decay products are characterized by high energy, so their detection becomes easier and, in addition.  the radioactivity  background is absent.}. If this optimistic scenario holds, the axion induced weak offers many advantages:
\begin{itemize}
    \item [i)] Many stable nuclei might become unstable. Provided that transitions $A(N,Z)(0^+)\rightarrow A(N-1,Z+1)(1^+)$
    become  energetically available with an axion energy of 5.5 MeV.
    One may think that  the targets involved in double beta decay \cite{ROP12} may undergo axion
    induced weak transitions. The relevant rates, however, are quite slow since the two neutrino double beta decay is faster and, in fact, it has been observed in many systems.\\ Better yet such transitions
    involving odd nuclei with many final states, some of which maybe populated even via the Fermi transition, become available.
    \item [ii)] In the case all ordinary  allowed beta decay transitions the end point energy will be shifted by
    5.5 MeV in the presence of axions.
    \item [iii)] Axion induced electron capture will exhibit a spectacular feature, which is the population  of an
     excited state in the final nucleus, not seen before, i.e. a 5.5 MeV above the highest one hitherto observed,
      if such a state with a similar structure exists.This can be a state not populated before, which will lead to
       a de-excitation  $\gamma$ ray of higher energy than hitherto  observed in the usual electron capture.
       The states already seen  can also be populated in the exotic process. Thus in this case
    \beq
    \frac{R^{E_x}_{0a,\beta}}{R^{E_x}_{0\beta}}=\frac{R_{0a,\beta}}{R_{0\beta}}\left (\frac{E_a+\Delta+m_e-b-E_x}{\Delta+m_e-b-E_x}\right )^2
    \eeq
    Where $\Delta$ is the available nuclear energy (Q-value), $b$ the binding energy of the electron and $E_x$ is
    the excitation energy of the final nuclear state. If the nucleus is chosen so that  $ \Delta+m_e-b-E_x$ is
    very small for some state, preferably that with the highest excitation energy, one may get a large enhancement in the rate.
    This is similar  to the well know nuclear excitation by neutrinos, photons,
     mesons and neutrons. The excited states decay mostly by gamma transitions and the weak
      decay branch is well know to be negligible ( of the order of $10^{-15}$).  The gamma  decays
      are non-negligible in the double beta decay and axion experiments.
The weak decays are known to be negligible in double beta decays
and axion search experiments, and thus are not considered there.
\end{itemize}

Before concluding this section we should mention that for axion induced weak processes the obtained rates are undetectable, especially if one uses the axion 5.5 MeV  Borexino flux. We are not going to elaborate here on this point, but  it will become apparent
after the  discussion of section \ref{sec:AxionInducedRates}.

\section{ The nuclear model}

For the evaluation of the cross sections of the processes discussed above,
an essential input is the nuclear matrix elements  for some targets of experimental
interest, preferably those that have been used in connection with dark matter searches.
Those are going to be obtained in the context of the large basis shell model.

\subsection{The target $^{127}$I}

In this work we study the experimentally relevant $^{127}$I for axion detection,
 The coherent contribution is expected to dominate in radiative axion capture, but this is pretty much independent
 of the details of the wave function.
 The spin-dependent (SD), however, depends on the details of the nuclear structure involved.

  The construction of the
nuclear wave functions needed for the evaluation of  matrix
elements entering in the corresponding scattering cross section is
accomplished in the framework of shell model. Specifically, the
calculation includes the valence space
$0g_{7/2},1d_{5/2},2s_{1/2},1d_{3/2},0h_{11/2}$ for protons and
neutrons on top of a $^{100}$Sn core.  The diagonalization of the
Hamiltonian was performed using the shell model code ANTOINE
\cite{Caurier2}. In order to make the calculations feasible, the
neutron configurations were restricted to allow at most
two-particle-two-hole excitations from the full
$0g_{7/2},1d_{5/2}$ shells. The number of excitations into the
orbitals $2s_{1/2},1d_{3/2},0h_{11/2}$ leads to a matrix dimension
of $7.3\times10^7$. A more extended configuration space has been
obtained in Ref.~\cite{Klos}. Proton many-body configurations were
not restricted due to small number of valence protons.   The mass
depended Bonn-C interaction \cite{Caurier1} was used as the
two-body interaction, while the single particle energies (SPEs):
$\epsilon_{0g_{9/2}}=-0.3$~MeV,
$\epsilon_{1d_{5/2}}=0.0$~MeV,$\epsilon_{2s_{1/2}}=1.3$~MeV,
$\epsilon_{1d_{3/2}}=1.5$~MeV, and $\epsilon_{0h_{11/2}}=1.9$~MeV
are taken from Ref. \cite{Ressell}.

\begin{figure}[htb]
    \begin{center}
        \includegraphics[width=0.48\textwidth]{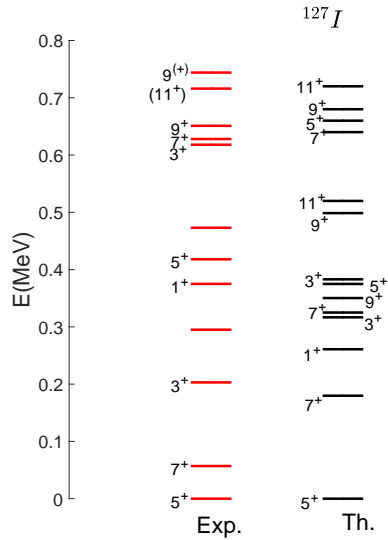}
        \caption{(Color on line) The low lying  experimental (left) \cite{Hashizume} and
            theoretical (right) energy spectrum of
            $^{127}$\hspace{1.0pt}I. The   levels in the figure are labeled by
            $2J$. Note that the spin and parity of the
            experimental states at 0.295 MeV and
           0.473 MeV are not known.
            }
        \label{fasma}
    \end{center}
\end{figure}

To test the quality of the structure calculations we compare the
energy spectrum and three known   magnetic moments of relevant
$^{127}$I isotope. Three magnetic moments are known for $^{127}$I.
Experimental and calculated magnetic moments are given in Table
\ref{tmm} while  in Fig.\ref{fasma} the corresponding energy
spectrum is presented.
 Most of our theoretical levels agree with the experimental ones to
about 100~keV.  An extension of the valence space to include
configurations that arise from excitations of one or more
particles probably might improve the agreement between theory and
experiment although there are experimental states that are still
uncertain in spin and parity assignment. Note that the spin and
parity of the low lying experimental states at 0.295 MeV and 0.473
MeV are not known. The identification of these states will be a
very useful test for our calculations.

\begin{table}[htb]
    \caption{Experimental \cite{Hashizume} and calculated magnetic moments in units of nuclear magneton $\mu_B(N)$.
        The magnetic moments are calculated using the standard g factors $g_{s,n}=-3.826$, $g_{s,p}=5.586$,
        $g_{l,n}=0$ and $g_{l,p}=1$ in nuclear magneton units.}
    \begin{center}
        \begin{tabular}{ccc}
            \hline
            state  &  Experiment    & Theory     \\
            \hline
            $5/2^+_{gs}$ & 2.81 & 2.92  \\
            $3/2^+_{1}$& 0.97 & 0.99  \\
            $7/2^+_{1}$& 2.54 & 1.72  \\
            \hline
        \end{tabular}
    \end{center}
    \label{tmm}
\end{table}

\begin{table}[p]
    \begin{center}
        \caption{   Isoscalar and isovector spin depended matrix elements
            of $^{127}$I from ground    ($5/2^{+}_{gs}$) to excited states  up
            to to the axion energy of 5.5 MeV, which are relevant for axion photoproduction. We also include the total nuclear
            matrix elements ME$^2(E_x)$, including that of the spin independent entering in the coherent mode,  which contain the factor
            $(1-\frac{E_x}{E_a})^3$ related to the photon energy, and the corresponding  cross sections.} \label{spinsI}
        \begin{tabular}{l|c|c|c|c|c }
            \hline
            $E_x$(MeV)          &$\Omega_0^2$ &$\Omega_1^2$& $\Omega_0\Omega_1$&ME$^2(E_x)$&$\sigma$$\times 10^{-48}$cm$^{2}$\\
            \hline \hline
             0    (\mbox{coherent})     &   &    &&4000&18714 \\

0.0000 &   1.0056 &  0.4140 & 0.6452  & 0.8042  & 3.217    \\
0.1800 &   0.1076 &  0.1243 & 0.1157  & 0.1326  & 0.480    \\
0.3120 &   0.0194 &  0.0583 & 0.0336  & 0.0417  & 0.140    \\
0.3150 &   0.0007 &  0.0212 & -0.0038  & 0.0064  & 0.022    \\
0.3750 &   0.0472 &  0.0314 & 0.0385  & 0.0452  & 0.146    \\
0.3830 &   0.0023 &  0.0194 & 0.0067  & 0.0105  & 0.034    \\
0.6490 &   0.0029 &  0.0198 & -0.0076  & 0.0054  & 0.015    \\
0.6600 &   0.0567 &  0.0481 & 0.0523  & 0.0603  & 0.164    \\
0.8160 &   0.0241 &  0.0996 & 0.0490  & 0.0643  & 0.159    \\
0.8340 &   0.0793 &  0.1188 & 0.0970  & 0.1120  & 0.274    \\
0.8610 &   0.0753 &  0.0135 & 0.0319  & 0.0477  & 0.114    \\
0.9290 &   0.0000 &  0.0001 & 0.0000  & 0.0001  & 0.000    \\
0.9780 &   0.3952 &  0.3163 & 0.3536  & 0.4094  & 0.910    \\
1.0440 &   0.0630 &  0.0523 & 0.0574  & 0.0663  & 0.141    \\
1.1570 &   0.0364 &  0.0167 & 0.0247  & 0.0302  & 0.060    \\
1.1700 &   0.0006 &  0.0006 & 0.0006  & 0.0007  & 0.001    \\
1.1930 &   0.0026 &  0.0562 & -0.0121  & 0.0166  & 0.032    \\
1.2790 &   0.0121 &  0.0112 & 0.0117  & 0.0134  & 0.024    \\
1.3660 &   0.0248 &  0.0093 & 0.0152  & 0.0192  & 0.033    \\
1.3910 &   0.0029 &  0.0007 & 0.0014  & 0.0020  & 0.003    \\
1.4780 &   0.0019 &  0.0859 & 0.0127  & 0.0365  & 0.057    \\
1.6380 &   0.0004 &  0.0035 & -0.0012  & 0.0010  & 0.001    \\
1.6780 &   0.0005 &  0.0100 & 0.0022  & 0.0046  & 0.006    \\
1.7610 &   0.0008 &  0.0055 & 0.0021  & 0.0031  & 0.004    \\
1.9370 &   0.0003 &  0.0019 & -0.0008  & 0.0005  & 0.001    \\
1.9650 &   0.0006 &  0.0047 & -0.0017  & 0.0013  & 0.001    \\
2.1040 &   0.0164 &  0.0202 & 0.0182  & 0.0209  & 0.020    \\
2.2370 &   0.0013 &  0.0097 & -0.0035  & 0.0027  & 0.002    \\
2.3140 &   0.0001 &  0.0002 & 0.0002  & 0.0002  & 0.000    \\
2.4280 &   0.0054 &  0.0036 & 0.0044  & 0.0052  & 0.004    \\
2.6030 &   0.0004 &  0.0002 & 0.0003  & 0.0004  & 0.000    \\
2.6650 &   0.0030 &  0.0001 & -0.0005  & 0.0011  & 0.001    \\
2.7420 &   0.0007 &  0.0002 & 0.0004  & 0.0005  & 0.000    \\
3.0070 &   0.0003 &  0.0005 & 0.0004  & 0.0004  & 0.000    \\
3.0620 &   0.0003 &  0.0000 & -0.0001  & 0.0001  & 0.000    \\
3.1740 &   0.0019 &  0.0009 & 0.0013  & 0.0016  & 0.000    \\
3.4440 &   0.0002 &  0.0001 & 0.0001  & 0.0002  & 0.000    \\
3.4890 &   0.0002 &  0.0000 & 0.0001  & 0.0001  & 0.000    \\
3.6020 &   0.0040 &  0.0009 & 0.0019  & 0.0027  & 0.000    \\
3.9230 &   0.0001 &  0.0004 & -0.0002  & 0.0001  & 0.000    \\
3.9680 &   0.0000 &  0.0002 & -0.0000  & 0.0001  & 0.000    \\
4.0600 &   0.0001 &  0.0002 & -0.0001  & 0.0001  & 0.000    \\
4.4340 &   0.0004 &  0.0000 & 0.0001  & 0.0002  & 0.000    \\
4.4560 &   0.0003 &  0.0000 & 0.0001  & 0.0001  & 0.000    \\
4.5550 &   0.0008 &  0.0002 & 0.0004  & 0.0005  & 0.000    \\
4.9430 &   0.0001 &  0.0000 & 0.0000  & 0.0001  & 0.000    \\
4.9570 &   0.0000 &  0.0000 & -0.0000  & 0.0000  & 0.000    \\
5.0370 &   0.0003 &  0.0000 & 0.0001  & 0.0002  & 0.000    \\
5.4340 &   0.0003 &  0.0000 & 0.0001  & 0.0002  & 0.000    \\
5.4730 &   0.0000 &  0.0000 & -0.0000  & 0.0000  & 0.000    \\
            \hline
            \hline
        \end{tabular}
    \end{center}
\end{table}

\begin{figure}[htb]
    \begin{center}
        \includegraphics[width=0.4\textwidth]{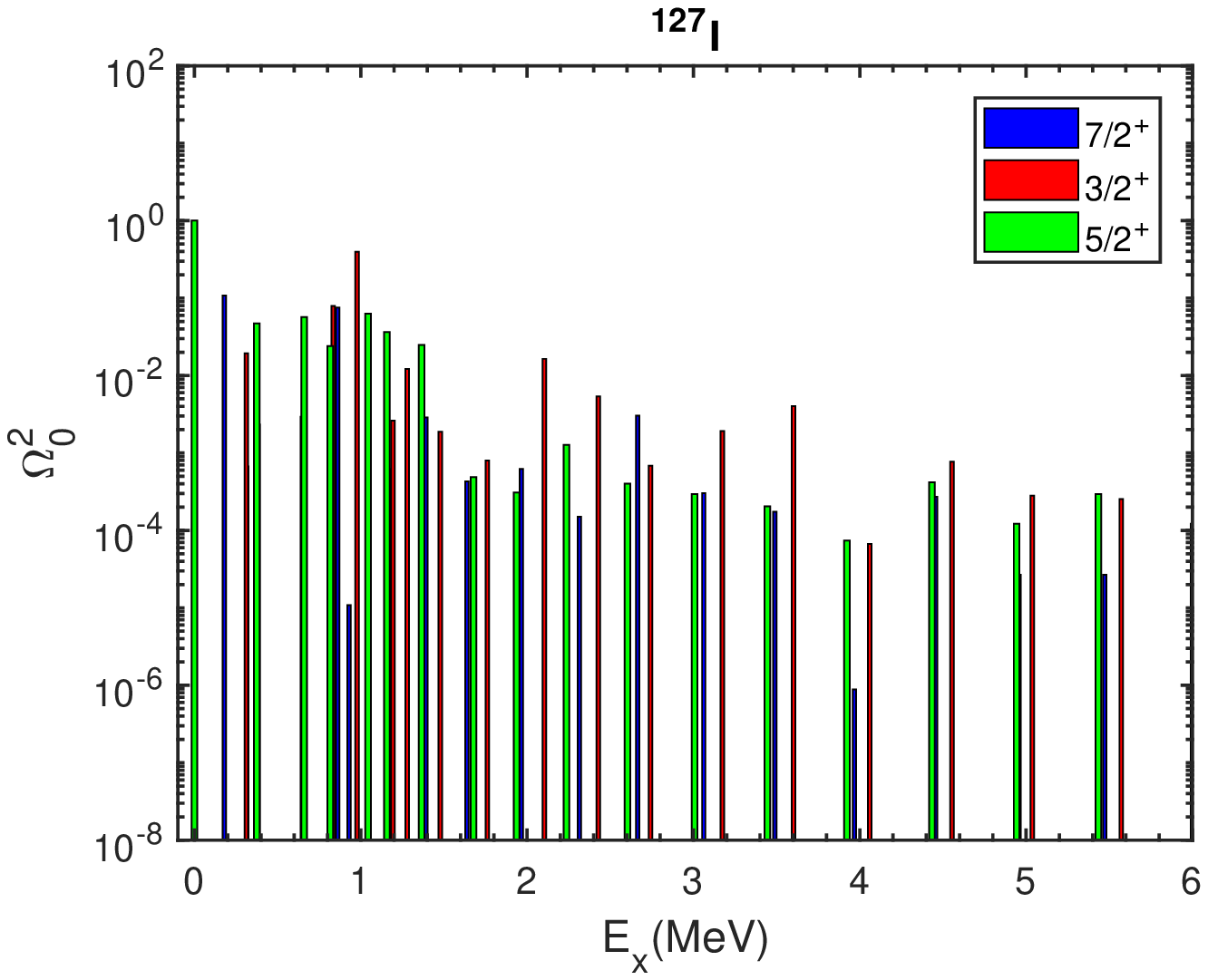}
        \includegraphics[width=0.4\textwidth]{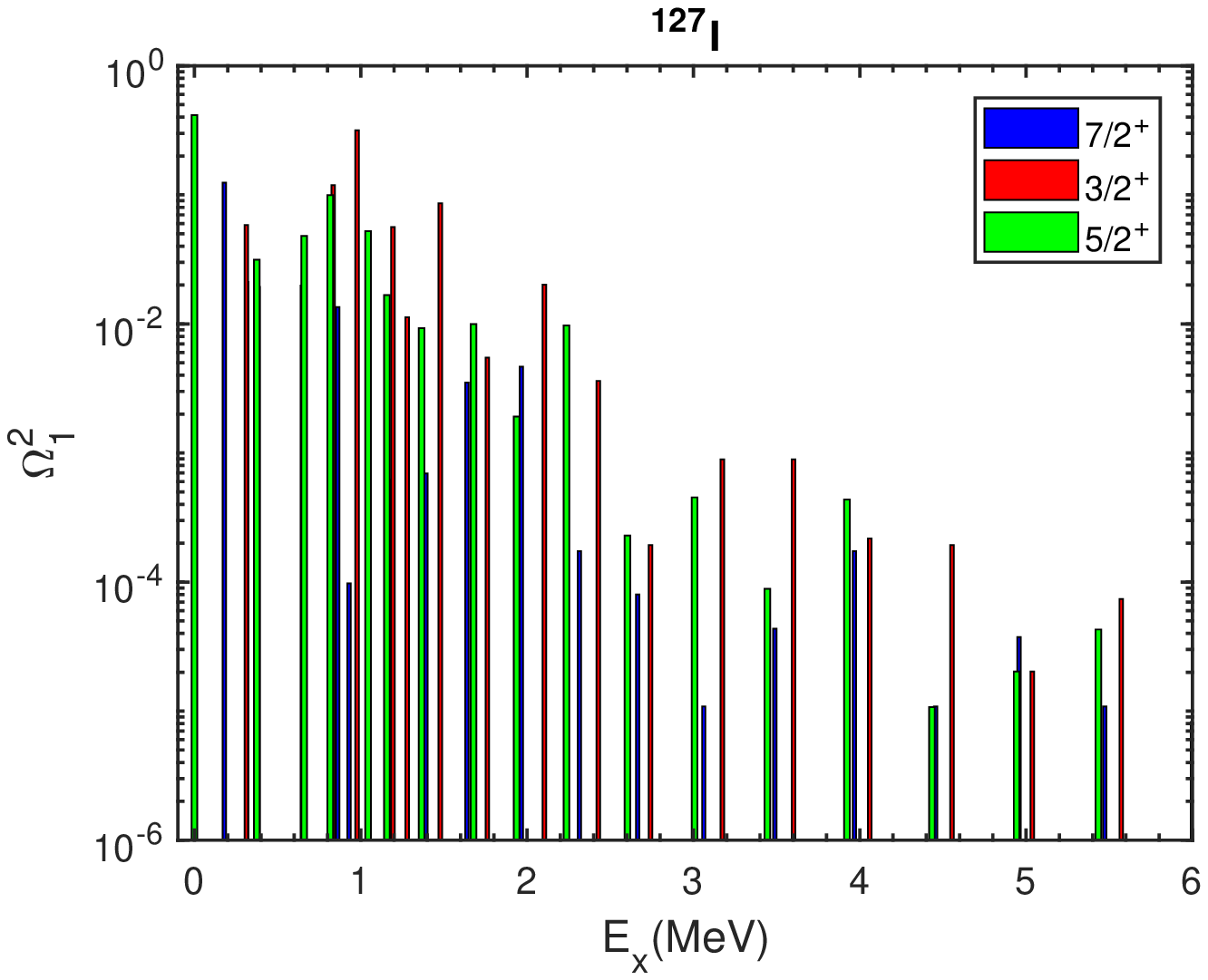}\hspace{1cm}
        \includegraphics[width=0.4\textwidth]{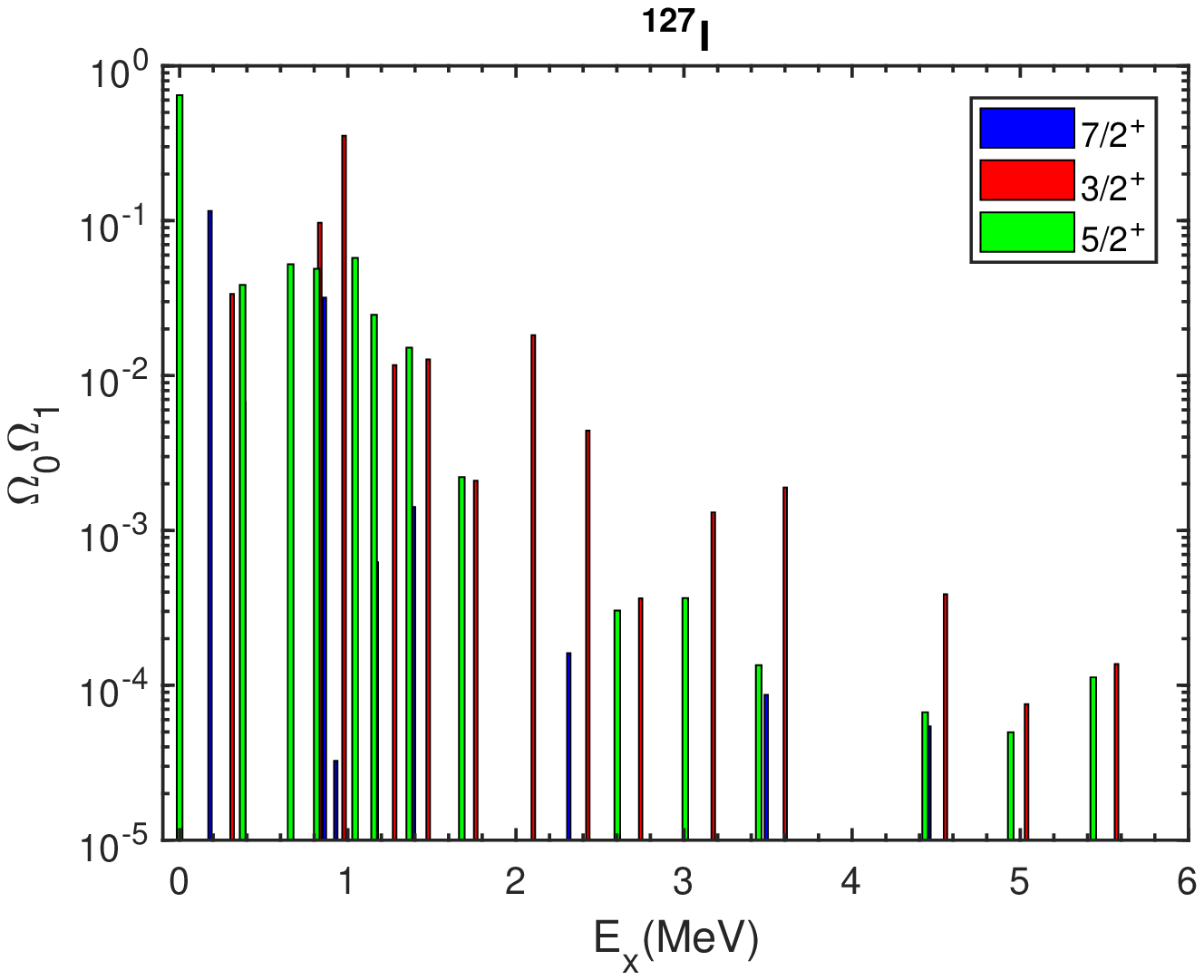}
        \caption{(Color on line) Isoscalar and isovector  distribution of
            $^{127}$I spin-reduced matrix elements from $ J_i={5/2}^+_{gs}$ to
            $J_f$ relevant for axion photoproduction. We should mention that
            the energy of the emitted photon is $E_{\gamma}=E_a-E_x$, i.e. the
            maximum photon energy corresponds to the gs to gs transition.}
        \label{omegas}
    \end{center}
\end{figure}

The isoscalar spin-reduced matrix elements
\beq
\Omega_{0}=\frac{\langle J_f||\sbf||J_i\rangle_{I=0}}{ \sqrt{{2J_i+1}}}\mbox{ (isoscalar) },
\Omega_{1}=\frac{ \langle
    J_f||\sbf||J_i\rangle_{I=1}}{ \sqrt{{2J_i+1}}}\mbox{ (isovector) }
\eeq from the ground state $ J_i=({5/2}^+)_{gs}$ to the  ground
state $ J_i=({5/2}^+)_{gs}$  and from the ground and to the final
$ J_f={3/2}^+,{5/2}^+$ and ${7/2}^+$ states   up to almost 6 MeV
have been computed. They are presented in Fig.\ref{omegas}. We
also summarize the obtained  results in table \ref{spinsI}. In
this table we also present the small nuclear matrix elements around the axion
energy $E_a=5.5$MeV, since the scale of the cross section involved,
see Eq.~(\ref{Eq:DeltaE2}), is orders of magnitude  larger than
the spin induced cross section involved in radiative axion
capture, Eq. (\ref{Eq:sgma0RadCap}). We also include the huge
matrix element involved in the coherent axion photoproduction.

\begin{figure}[htb]
    \begin{center}
        \includegraphics[width=0.4\textwidth]{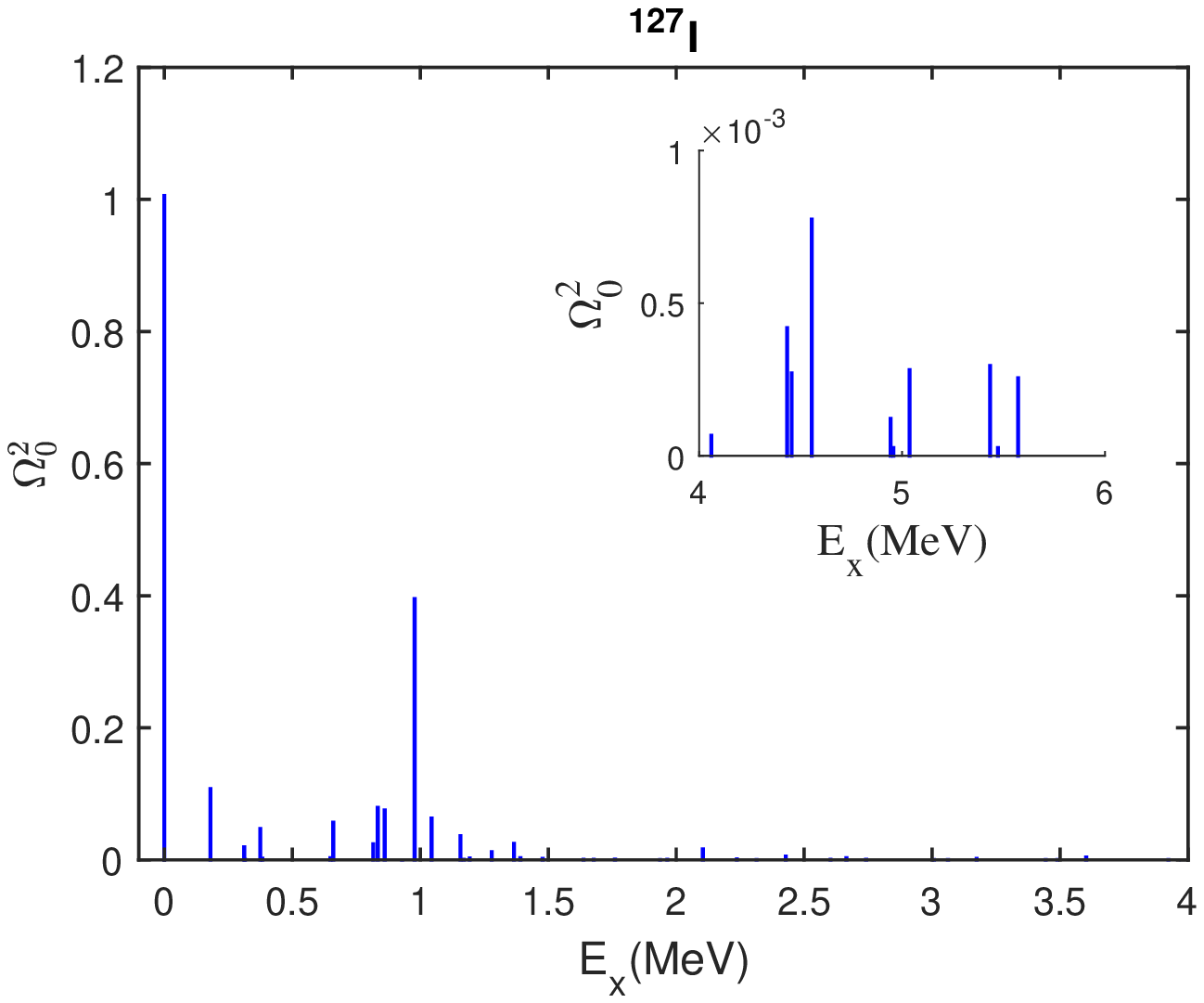}
        \includegraphics[width=0.4\textwidth]{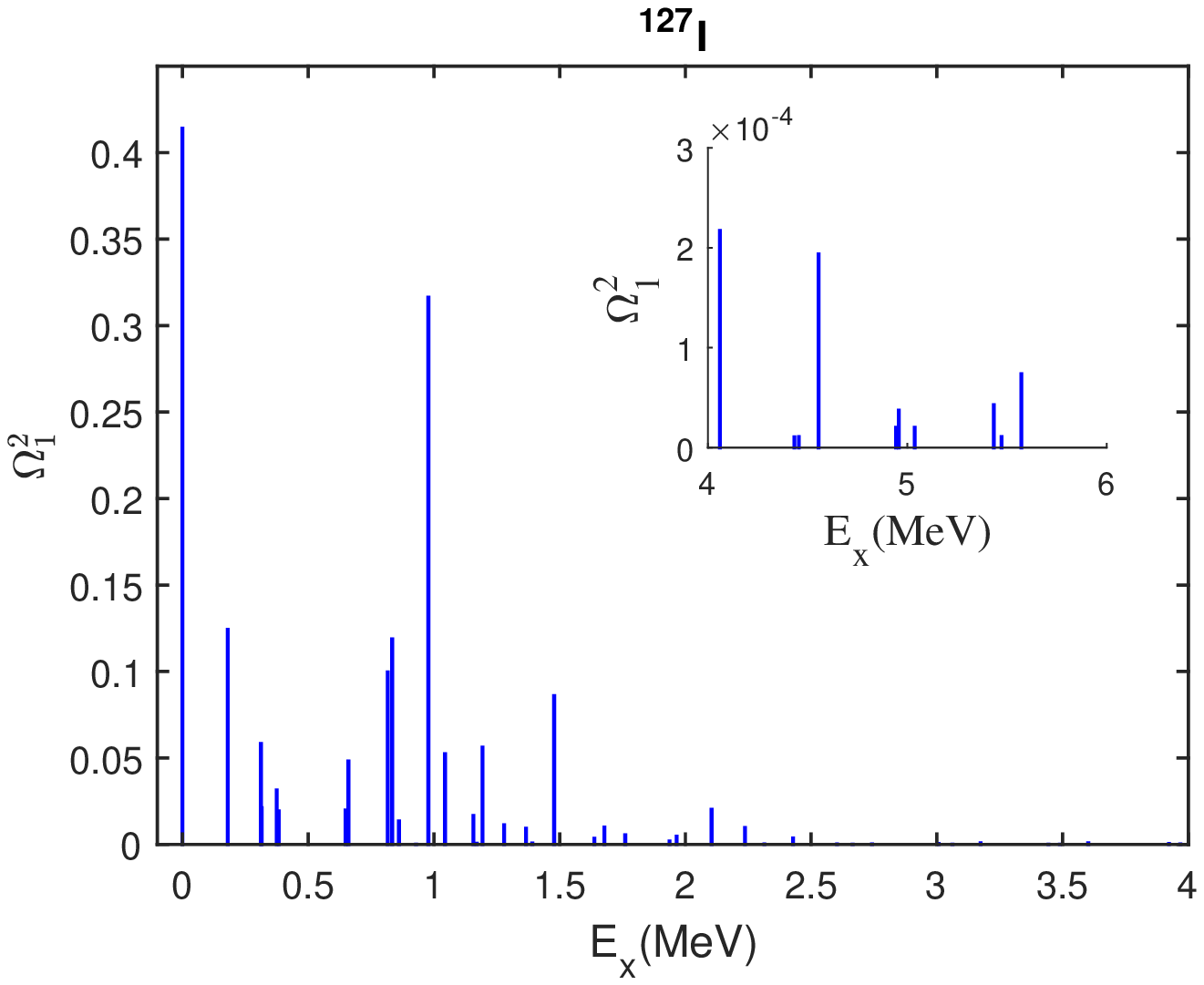}\hspace{1cm}
        \includegraphics[width=0.4\textwidth]{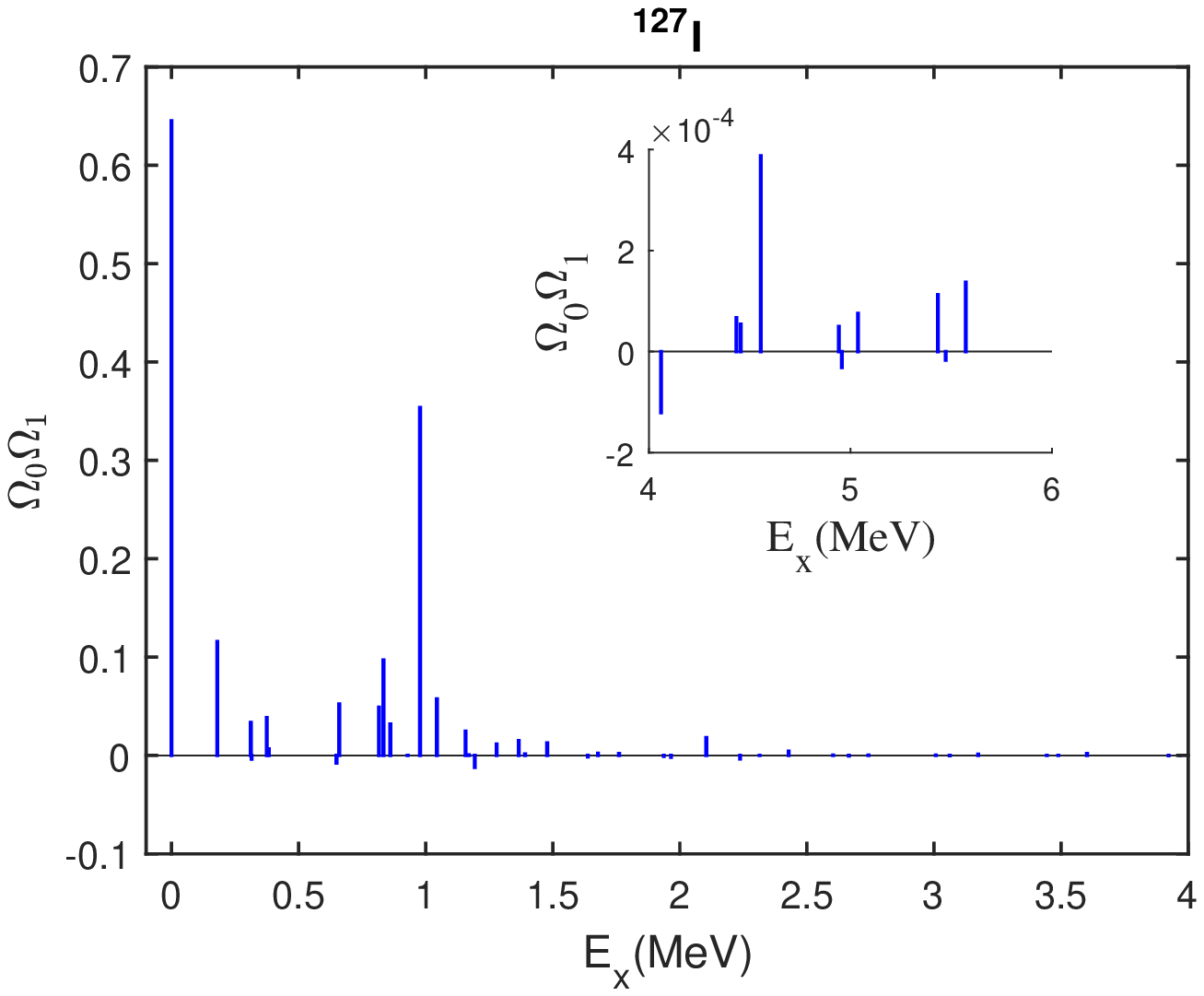}
        \caption{(Color on line) Isoscalar and isovector spin matrix elements
            up to 6 MeV relevant
            for axion absorption are presented. Note that in the range of 4-6 MeV, in order to make the matrix elements visible, we exhibit them in a separate small window, where the small scale is indicated.}
        \label{omega}
    \end{center}
\end{figure}

\begin{table}[htb]
    \begin{center}
        \vspace{5pt} \caption{   Isoscalar and isovector spin depended
            matrix elements of $^{23}$Na from ground    ($3/2^{+}_{gs}$) to the various
            excited states entering the axion induced photoproduction. We also include the total nuclear
            matrix elements ME$^2(E_x)$, including the spin independent one appearing in the coherent mode, which contain the factor
            $(1-\frac{E_x}{E_a})^3$ related to the photon energy, and the
            relevant cross sections as well. Note that states up to the axion energy of 5.5 MeV are relevant.} \label{spinsNa}
        \renewcommand{\tabcolsep}{0.5pc} 
        \renewcommand{\arraystretch}{1.0} 
        \begin{tabular}{l|c|c|c|c|c}
            \hline
$E_x$(MeV)          &$\Omega_0^2$ &$\Omega_1^2$& $\Omega_0\Omega_1$&ME$^2(E_x)$&$\sigma$$\times 10^{-48}$cm$^{2}$\\
            \hline
             0    (\mbox{coherent})     &   &    &&200&  791\\
   0.0000 &   0.4756 &  0.3450 & 0.4051  & 0.4721  & 1.889      \\
0.4120 &   0.2024 &  0.2589 & 0.2289  & 0.2628  & 0.832      \\
2.3240 &   0.1868 &  0.1286 & 0.1550  & 0.1814  & 0.140      \\
2.7470 &   0.4689 &  0.3665 & 0.4145  & 0.4807  & 0.241      \\
3.8930 &   0.0934 &  0.0928 & 0.0931  & 0.1069  & 0.011      \\
4.3110 &   0.5066 &  0.4589 & 0.4821  & 0.5551  & 0.022      \\
5.2460 &   0.0931 &  0.1184 & 0.1050  & 0.1205  & 0.000      \\
5.7460 &   0.0474 &  0.0708 & 0.0579  & 0.0668  & -        \\
5.9760 &   0.0105 &  0.0046 & 0.0070  & 0.0086  & -       \\
            \hline
        \end{tabular}
    \end{center}
\end{table}

\subsection{The target $^{23}$Na}

The valence space of    $^{23}$Na  is the $sd$ shell, which
comprises the $0d_{5/2}$, $1s_{1/2}$, and $0d_{3/2}$ orbitals,
with a $^{16}$O core. Full calculations for this isotope have
previously been  performed  in the work~\cite{Diva00}.

\begin{figure}[htb]
    \begin{center}
        \includegraphics[width=0.4\textwidth]{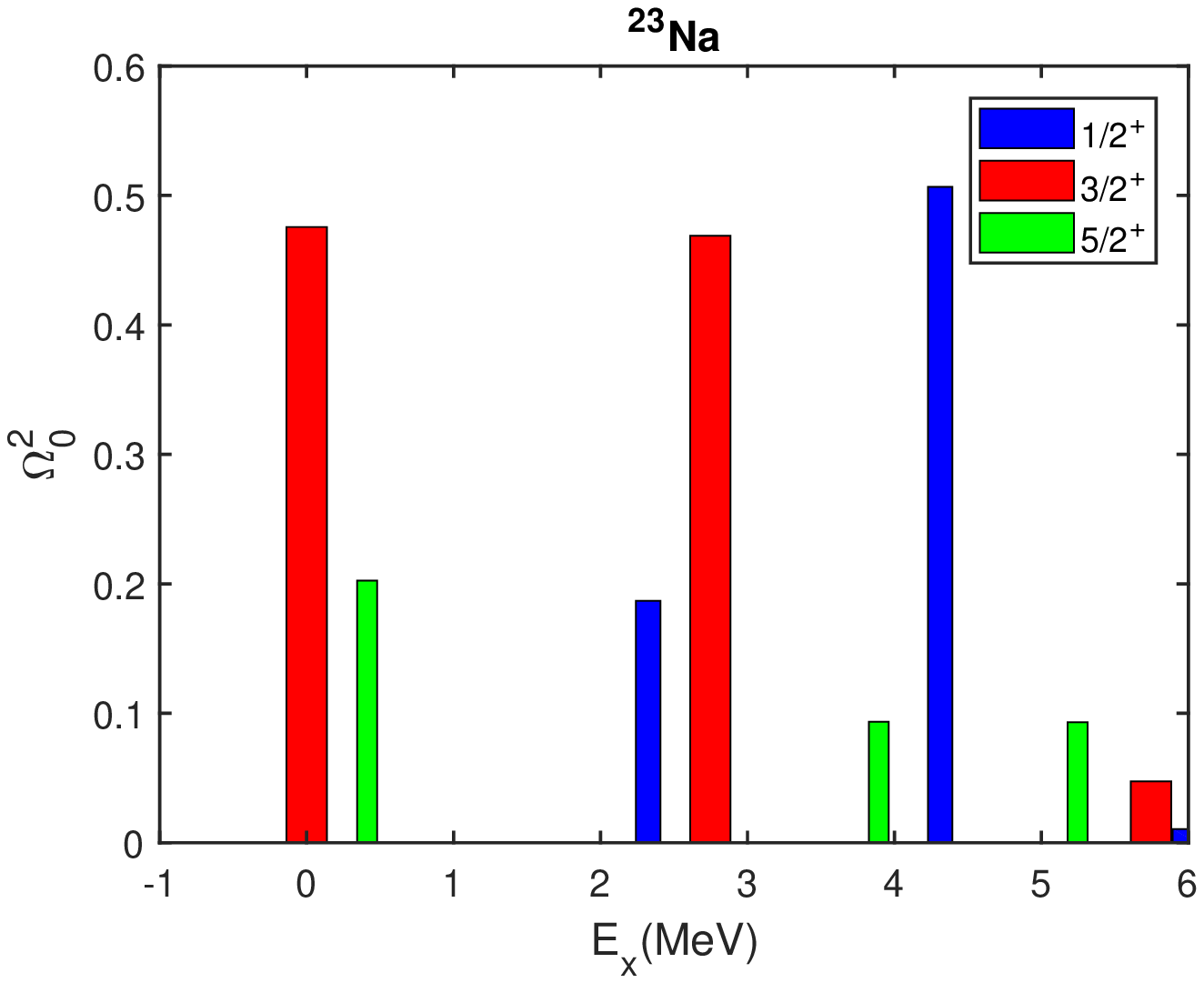}
        \includegraphics[width=0.4\textwidth]{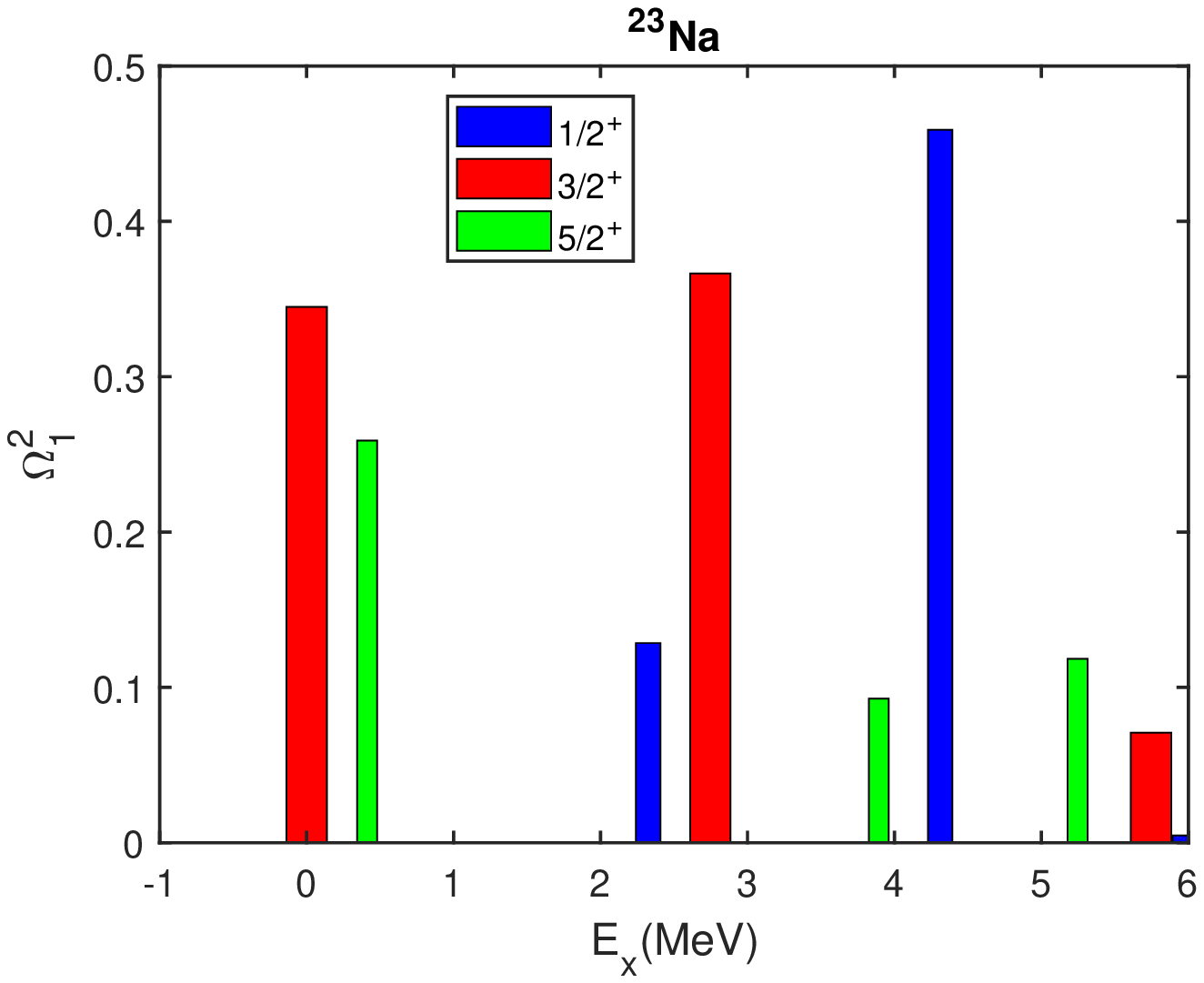}\hspace{1cm}
        \includegraphics[width=0.4\textwidth]{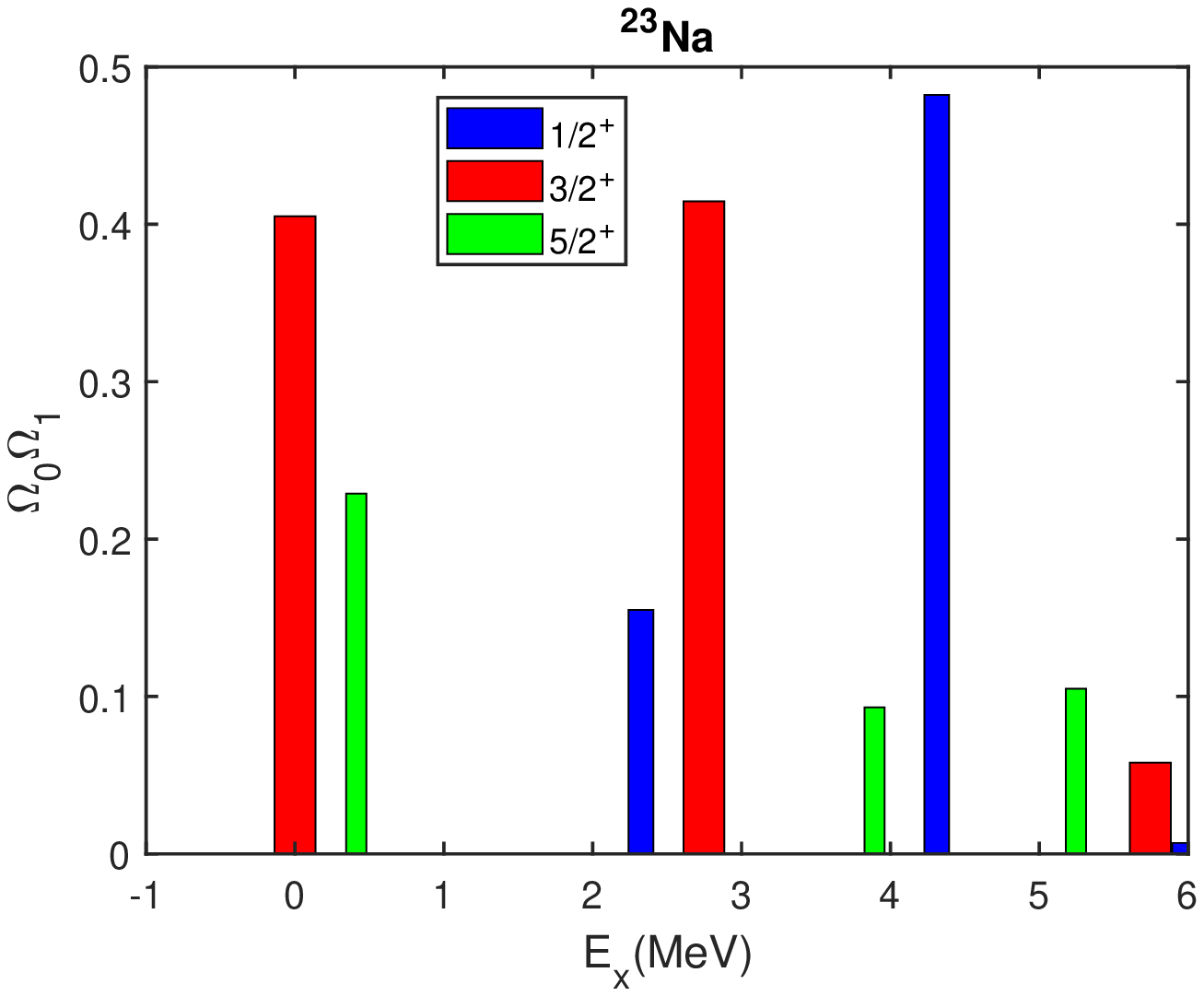}
        \caption{ Isoscalar and isovector  distribution of $^{23}$Na
            spin-reduced matrix elements from $ J_i={3/2}^+_{gs}$ to $J_f$
            relevant for axion photoproduction. We should mention that the
            energy of the emitted photon is $E_{\gamma}=E_a-E_x$, i.e. the
            maximum photon energy corresponds to the gs to gs transition. }
        \label{omegaNa}
    \end{center}
\end{figure}

As expected  from weak interactions the spin induced rates at low excitation energies are
small compared to some canonical value (sum rule). Indeed
\barr
\frac{1}{2 J_i+1} \sum_f \langle J_f||{ \bf O}||J_i\rangle^2&=&\sum_{M_i,M_,q,f} (-1)^q \langle J_fM_f||{ \bf O}_q|J_iM_i\rangle^2 = \sum_{M_i,M_,q,f} (-1)^q\langle J_i M_i|{ \bf O}_{-q}|J_f M_f\rangle|\langle J_f M_f|{ \bf O}_q|J_i M_i\rangle\nonumber\\
&=&\sum_{q}(-1)^q\langle J_i M_i|{ \bf O}_{-q}{ \bf O}_{q}| J_i M_i\rangle=\langle J_i M_i|{ \bf O}\cdot{ \bf O} |J_i M_i\rangle
\label{Eq:sumrule}
\earr
with
\beq
{ \bf O}=\sum_{k}\left (g^0_{aN}+g^3_{aN}\tau_3(k) \right )\sbf(k)
\eeq
As a result Eq. (\ref{Eq:sumrule}) produces an one body part which
is
\beq
\left (\left (g^0_{aN}\right )^2+\left (g^3_{aN}\right )^2 \right )A+2 g^0_{aN}g^3_{aN}(2 Z- A),
\label{Eq:cohsumofs}
\eeq
and a two body
part which is $2V$ with
\beq
V=\langle J_i M_i|\sum_{k \ell} \left
((g^0_{aN}+g^3_{aN}\tau_3(k) )\sbf{(k)}\cdot
(g^0_{aN}+g^3_{aN}\tau_3(\ell)) \right )\sbf{(\ell)}| J_i
M_i\rangle.
\eeq
Since $V$ is expected to be much smaller than   that given by Eq, (\ref{Eq:cohsumofs}), we find to a ggod approximation
\beq
S\approx \left (\left (g^0_{aN}\right )^2+ \left (g^3_{aN}\right )^2 \right )A+2 g^0_{aN}g^3_{aN}(2 Z- A)
\eeq
We do not know the energy where this strength is located, but is expected to be in the higher part of the spectrum.

In the special case of interest to us using the couplings given by Eq. (\ref{Eq:effANucCoup}) and model F of (\ref{Eq:bestF}) we find
\barr
S(I)&=&(0.083^2+0.192^2)\times 127+2\times 0.083 \times 0.192 \times(-11)=5.1,\nonumber\\
S(Na)&=&(0.083^2+0.192^2)\times 23+2\times 0.083 \times 0.192 \times(-1)=1.0
\earr

\section{Event rates}

Noting that the oncoming beam of axions is monochromatic, the
event rate is given by: \beq R=\sigma \Phi(E_a) N_A
\label{Eq:Fe57SFv}\eeq where $\sigma$ is cross section obtained
above,  $ N_A$ the number of nuclei in the target and  $\Phi(E_a)$
the oncoming  axion flux.
We will consider two cases:\\
i) the 14.4 keV Fe*$^{57}$ flux
\beq \Phi(E_a)=0.7 \times 10^9
\left (\frac{10^{7}\mbox{GeV}}{f_a}\right )^2\mbox{cm}^{-2}\mbox{s}^{-1}
\mbox{ (14.4 keV solar axion)} \label{Eq:Fe57SF}
\eeq
The flux has
been obtained using the calculated
 matrix elements \cite{SolFlux20} for the $^{57}$Fe* de-excitation
 and realistic axion nucleon coupling constants obtained in a variety of particle models.

ii) the 5.5 MeV He-3 flux\\
The flux of the 5.5 MeV solar axions is obtained   under the assumption
that the axion production is dominated by the isovector transition and
 an axion mass much smaller than its energy. The standard formula for
 axion to photon production rate  \cite{Peccei96,SolFlux20} can be cast in the form:
\beq
\frac{\Gamma_{a}}{\Gamma_{\gamma}}\approx 8\times 10^{-13}\left ({g^3_{aN}}\right )^2\left (\frac{m_a}{5 \mbox{eV}}\right )^2
\eeq
From this  one can obtain the axion flux, if the photonic rate for the
 process of Eq. (\ref{Eq.SolarFlux}) is known. In the Standard Solar Model (SSM) this constitutes the
second stage of the pp solar fusion chain, with the first stage
provided by the two reactions $p + p \rightarrow d + e^+ + \nu_e$
and $p + p + e^- \rightarrow d + \nu_e $. As the deuterons
produced via this first stage capture protons within $\tau=6$s,
the axion flux resulting from (\ref{Eq.SolarFlux}) can be
expressed in terms of the known pp neutrino flux
\cite{Borexino12,BHHousLi20}:
$$\Phi_{\nu pp}=6.0\times  10^{10}\mbox{ cm }^{-2} \mbox{s}^{-1} \mbox{ (solar pp neutrino flux)} $$
 The  extracted flux  from the Borexino experiment \cite{Borexino12}   for the 5.5 MeV axions is given by
\beq
 \Phi=2.45 \times 10^{-5}
\mbox{cm}^{-1}\mbox{s}^{-1}(m_a/\mbox{1 eV})^2=6.1 \times
10^{-4}\mbox{cm}^{-2}\mbox{s}^{-1}(m_a/\mbox{5 eV})^2 =1.86\times 10^4\mbox{cm}^{-2}\mbox{y}^{-1}(m_a/\mbox{5\ eV})^2
\label{Eq:NewBorFlux}
 \eeq
with the last re-scaled value obtained for $m_a=$5eV. This is many orders of
magnitude smaller than that of Eq. (\ref{Eq:Fe57SF}).

We will consider  $N_A=4\times 10^{27}$, which is close to 1ton of
NaI in the target.

\section{Some results for nuclear transitions}

NaI seems to be a ideal material for the detection of $\gamma$
rays. Such a target has already been employed by the DAMA
experiment in dark matter searches, see e.g.\cite{DAMA99}, and
later in the DAMA/LIBRA collaboration \cite{DAMALIBRA08}.

\subsection{Axion photoproduction}

Let us begin with the axion photoproduction.\\

A) We will first examine  the component $^{127}$I of the target. We will consider two cases \\
a) Elastic scattering  contributions. \\
The first term is the coherent contribution which amounts to
\beq
\mbox{ME}^2_{\mbox{coh}}=\frac{2}{3}(0.781\times 127 + 0.734 \times(-21))^2\approx 4 \times 10^{3}
\eeq
The second term involves the gs$\rightarrow$gs spin induced transition obtained from  Fig \ref{omega} we find:
\barr \mbox{ME}^2_{\mbox{gs}}=\frac{2}{3}\left
((G^0_{aN}\Omega_0)^2+(G^3_{aN}\Omega_1)^2+
G^0_{aN}G^3_{aN}\Omega_0\Omega_1 \right)=&  \\ \nonumber
\frac{2}{3}\left
((G^0_{aN})^21.005+(G^3_{aN})^20.414+
G^0_{aN}G^3_{aN}0.645 \right)= 0.804
\earr

b) Similarly for the transition to the excited state at 0.978  MeV we obtain:
\barr \mbox{ME}^2_{1}=\frac{2}{3}\left
((G^0_{aN}\Omega_0)^2+(G^3_{aN}\Omega_1)^2+
G^0_{aN}G^3_{aN}\Omega_0\Omega_1 \right)=& \\ \nonumber
\frac{2}{3}\left ((G^0_{aN})^2 0.395+(G^3_{aN})^2 0.316+
G^0_{aN}G^3_{aN}0.354 \right)= 0.409
\earr

Using now Eq.  (\ref{Eq:Fe57SFv}) we obtain and considering only
the above  the dominant contributions we obtain  the event rate
\barr
E_{\gamma}=5.5 \mbox{MeV}\leftrightarrow R_{\mbox{\small coh}}&=&2.9 \times 10^{-16} \times 4 \times 10^{3}\approx 1.2\times 10^{-12}  \mbox{(t-y)}^{-1}\nonumber\\
E_{\gamma}=5.5 \mbox{MeV}\leftrightarrow R\Big({\frac{5}{2}}^+\Big)_{gs}&=&2.9 \times 10^{-16} \times 0.804\approx 2.4\times 10^{-16}\mbox{(t-y)}^{-1},\nonumber\\
E_{\gamma}=4.5 \mbox{MeV}\leftrightarrow R\Big({\frac{3}{2}}^+\Big)_4&=&2.9 \times 10^{-16} \times 0.409\times \left (1-\frac{E_x}{E_a} \right )^3=2.9 \times 10^{-16} \times 0.409\times 0.55 \approx 0.6\times 10^{-16}\mbox{(t-y)}^{-1}\nonumber\\
\label{Eq:ResPhotoI} \earr

A state by state evaluation of the rates can be similarly  obtained from the cross sections is given in Table
\ref{spinsI}.

B) The component $^{23}$Na of the target. \\The obtained rates can
also be obtained from the cross sections   summarized in Table
\ref{spinsNa}. Using these data we find for the dominant
contribution
$$R_{\mbox{\small coh}}=6\times 10^{-14} \mbox{(t-y)}^{-1}, $$
expected due to the smaller A involved.

It is clear that the coherent  ME is very large and dominates. We do not know of any
 other process that exhibits such a coherence. In pion photoproduction see e.g. \cite{EricRho72}  and \cite{JDV75}
 one encounters a spin independent contribution, but this does not lead to coherence since the pion is negatively
 charged and stop in an atomic orbit before it is absorbed by the nucleus. Anyway a change of the nucleus is involved.
 The process $(\gamma\pi^0)$ is somewhat different, but it leads to
 coherence. The coherent scattering  of the axion is the same as the well-known coherent scattering
of low-energy (large wave-lengths) neutrons, photons, electrons,
neutrinos  and others. They are traditionally called as elastic
scattering.

In any case we believe that the coherence is there, but, in spite of the fact that the axion photoproduction
 is favored by the nuclear physics, the obtained rate is not very large. This can be traced to the small axion mass
 which makes the relevant coupling small. The above rates were obtained with $4\times 10^{27}$ particles in the target,
 which amounts  to 1ton of NaI. Close to such amounts of material have in fact been used in
  WIMP (weakly interacting massive particles) dark matter searches \cite{DAMALIBRA08,DAMALIB18,KCs12}.

\subsection{Axion induced nuclear excitation}
\label{sec:AxionInducedRates}
A) We will begin with the target $^{127}$I\\
In this case we consider the excitation of the nucleus which can
subsequently de-excite by $\gamma$ emission. In this case, since
the axion source is monochromatic, only states around   $E_a$ can
be excited. The nuclear matrix elements can be obtained form those
of Fig. \ref{omega} Thus we obtain
$$\left(g^0_{aN} \right) ^2(0.8+0.3)\times 10^{-4}+\left(g^3_{aN} \right) ^2 4 \times 10^{-4}+ \left(g^0_{aN} g^3_{aN}\right) 4 \times 10^{-4}=2.2 \times 10^{-5}.$$
Thus for $\kappa=0.1$ in Eq. (\ref{Eq:DeltaE3}) and  using Eqs
(\ref{Eq:Fe57SFv}) and (\ref{Eq:NewBorFlux}) we obtain \beq R_5=
1.86 \times 10^{4} \times \frac{1}{0.2}\times6.28\times
10^{-41}\times 2.2 \times 10^{-5}\times
4\times10^{27}\mbox{y}^{-1}=5 \times 10^{-13}\mbox{(t-y)}^{-1}
\label{Eq:RateR5I} \eeq
This time the scale of the cross section is favorable, but the nuclear matrix elements are suppressed.\\

B) We will continue  with the target $^{23}$Na\\

In this case the last states of Table  \ref{spinsNa} are relevant.
Proceeding as in the previous case we obtain: \beq
(6.4,\,1.5,\,0.9,\,0.07)\times10^{-10} \mbox{(t-y)}^{-1} \mbox{in
that order} \label{Eq:RateR5Na} \eeq

Unlike the case of $^{127}$I, the  above rates for $^{23}$Na appear
larger, but still small.

In view of the very small 5.5 MeV axion flux for $m_a=$5 eV axions, as given by eq. (\ref{Eq:NewBorFlux}),
axion motivated  nuclear physics experiments look doomed.

\begin{figure}[htb]
    \begin{center}
        \subfloat[]
       {
        \includegraphics[width=0.4\textwidth]{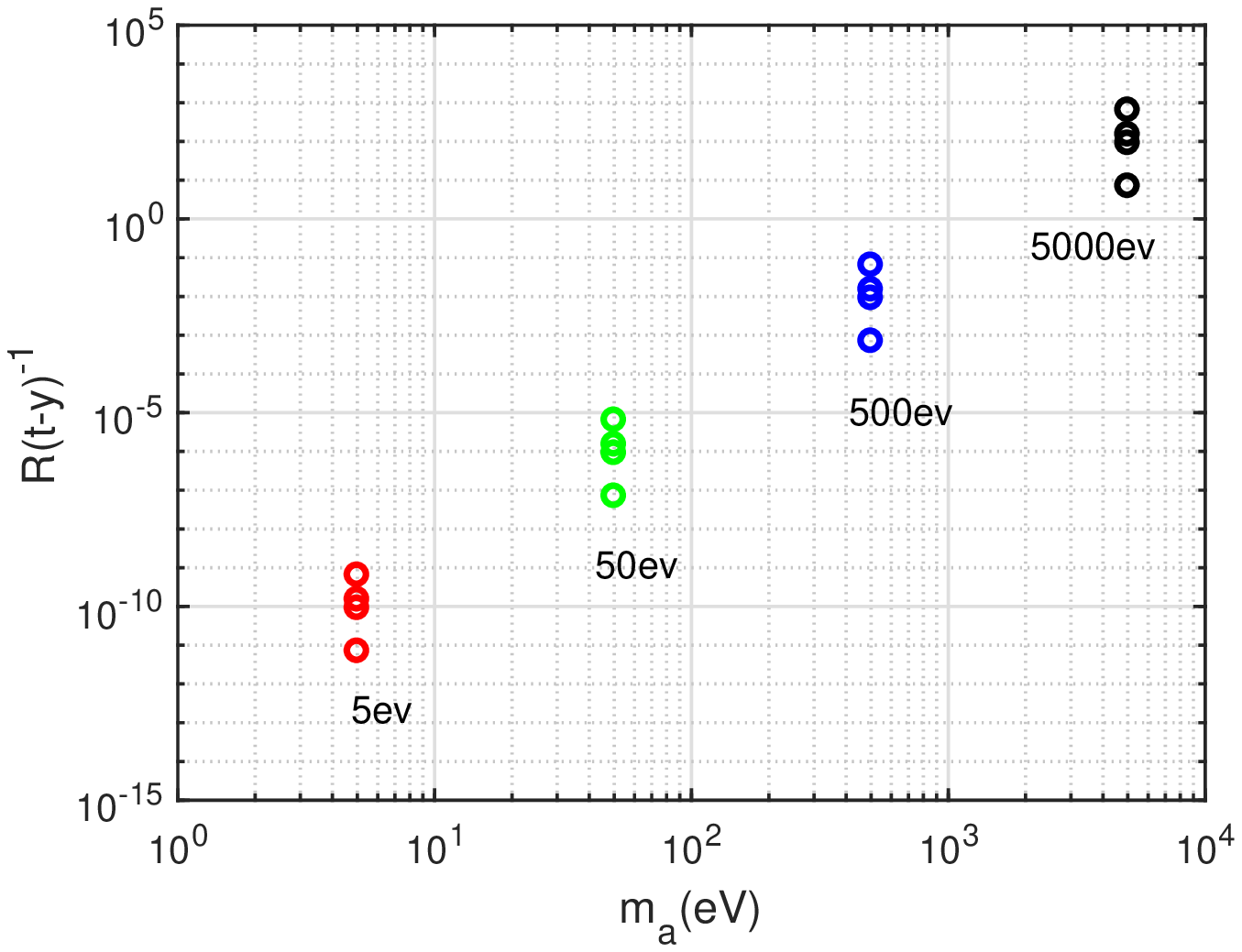}
       }
   \subfloat[]
   {
    \includegraphics[width=0.4\textwidth]{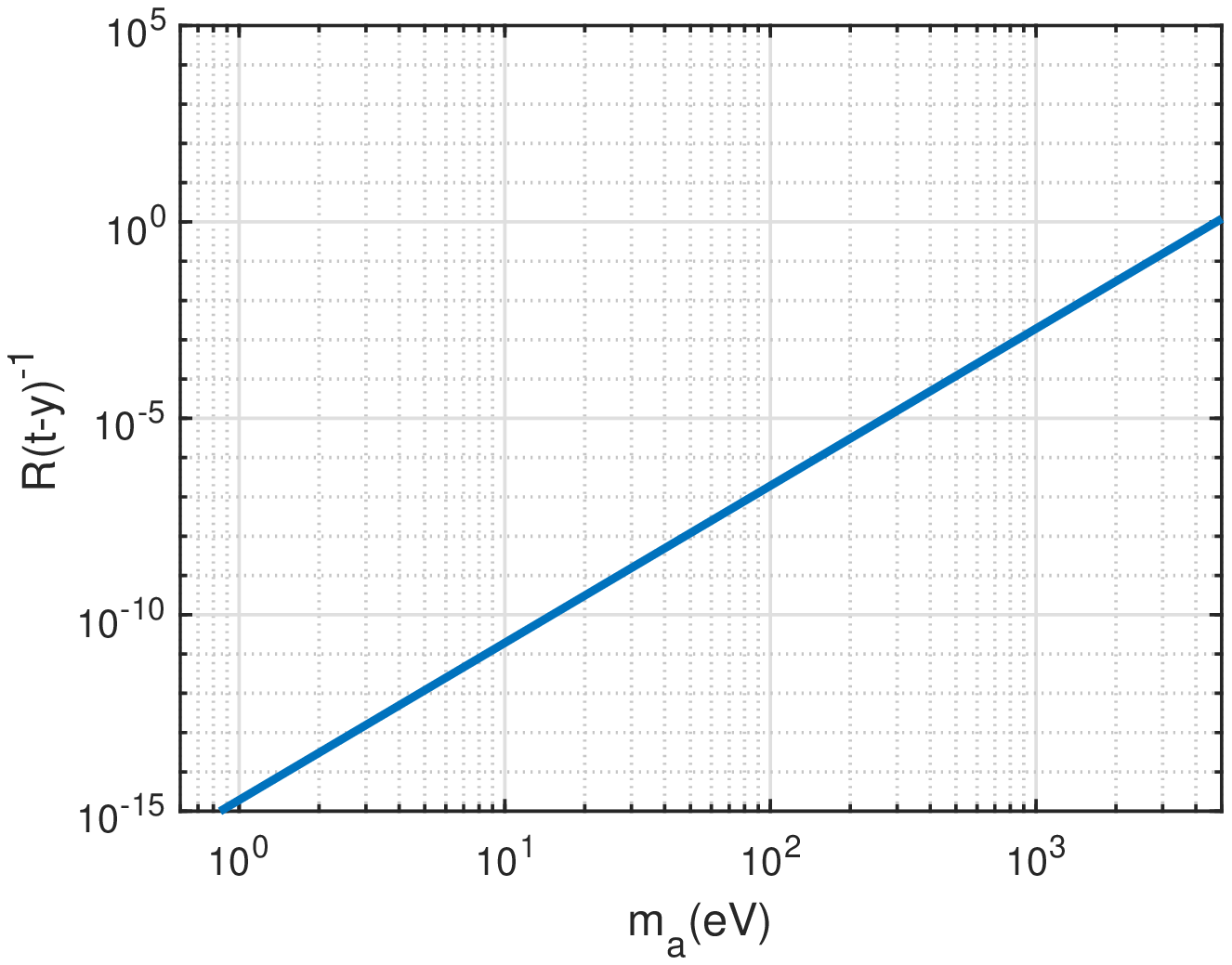}
   }
        \caption{(Color on line)
        (a)Event rates for the last four states 4.31MeV, 5.24MeV, 5.74MeV, and
        5.97MeV of $^{23}$Na
       (see Table \ref{spinsNa}) versus different axion masses.(b) The rate for
       axion photoproduction on $^{127}$I.
        As expected this is quite a bit smaller than in (a),
        due to the fact that the axion energy of 5.5 MeV is much
         smaller than the nucleon mass. Note, however, that in this
          case we have the extra signature  of the produced photon.
           Since the  rate is dominated by the coherent mode,
           the photon energy is approximately equal to the axion energy.}
        \label{rvma}
    \end{center}
\end{figure}

As we have seen, however, the axion mass is not known. The
obtained rates scale with $m^2_a \times m^2_a$, one factor coming
from the axion flux and the other of the axion-nucleus cross
section.  The obtained rates as a function of the axion mass for two typical cases are exhibited in Fig. \ref{rvma}.
For a larger mass, e.g. 5 keV, the expected  rates become:\\
(a) For the last four states 4.31MeV, 5.24MeV, 5.74MeV, and
5.97MeV  of $^{23}$Na (see Table \ref{spinsNa}) as follows
\beq R=(6.4,\,1.5,\,0.9,\,0.07)\times10^{2} \mbox{(t-y)}^{-1}
\label{Eq;NucBorRates} \eeq In the case of the expressions
(\ref{Eq:ResPhotoI}), (\ref{Eq:RateR5I}) and  (\ref{Eq:RateR5Na})
respectively.\\
The rates for $^{23}$Na excitation are higher than those for
$^{127}$I, since it so happens that in the former case states with
larger nuclear matrix element happened to be around 5.5 MeV.

We remind the reader that only excited states with energies in a narrow window around 5.5 MeV can contribute to the process. The structure of all the other states, especially of the low lying ones, whose energies  are experimentally known and can be compared to the calculated ones, is immaterial.

(b) For the axion photoproduction rate, dominated by the coherent
mode of $^{127}$I, as follows: \beq R\approx R_{coh}=1.2
\mbox{(t-y)}^{-1} \label{Eq;NucPhoProdRates} \eeq
This rate is expected to be the similar to that of nuclear targets with (A,Z) close to that of $^{127}$I.\\
As expected this rate is quite a bit smaller than in (a), due to the fact that the axion energy of 5.5 MeV
is much smaller than the nucleon mass. Note, however, that in this case we have the extra signature
of the produced photon. Since the  rate is dominated by the coherent mode,  the photon energy
is approximately equal to the axion energy.\\
Note that if only the total $\gamma$-ray energy is measured, one cannot distinguish between (a) and (b).

It seems that axion masses in the keV region are not allowed by
astrophysical and cosmological axion limits, see e.g. Fig. 3 of
the review by Raffelt \cite{GRaffelt08}. So, even though such
limits are somewhat model dependent,  we have to concede that
axion masses in the keV region are not realistic and high
detection rates as those mentioned above in Eqs
(\ref{Eq;NucBorRates}) and (\ref{Eq;NucPhoProdRates}) are not
expected.

In addition to the 5.5 MeV solar axions employed here, other solar
axion sources of quite a bit lower energies  can be employed in
nuclear physics experiments, like  the  477.6 keV  $^{7}$Li solar
axion. A search  for such axions  has been carried out
\cite{Krcmar1,Derbin1,Belli1} by using the resonant axion
absorption process. No such axions have been detected, but the
data  collected  allowed the authors to set a limit on the
$^{7}$Li solar axion mass of 13.9 keV at 90\% C.L. \cite{Belli1}.
It is worth mentioning that a LiF powder of 243 g measured during
722 h was employed, i.e. $2.0\times 10^{-5}$ (t-y). This
corresponds to a detector sensitivity  of about $1.5\times 10^{6}$
(t-y)$^{-1}$. It is obvious that this sensitivity needs to be
improved by many orders of magnitude to reach the level of, e.g.,
$10^2$(t-y)$^{-1}$ expected for 5 eV axions and relevant  coupling
to the electron of $g_{ae}=024$, see section \ref{sec:eresults}
below. This proposal  is allowed and interesting, expected  to be
considered in future experiments. We are not going to employ the
13.9 keV axions in this work, since  the authors themselves
classify  their work as " Preliminary results and feasibility
studies".

If, instead,  we use lower masses, e.g.  the value of  $m_a=0.25$keV, a model dependent
limit extracted by the CUORE experiment \cite{CUOREColl13}, see the discussion
 following Eq. (\ref{Eq:CUOREbounds}) below, we find
\beq
 R=(6.4,\,1.5,\,0.9,\,0.07)\times 6.25\times 10^{-4} \mbox{(t-y)}^{-1},\,R_{coh}=7.5 \times 10^{-6} \mbox{(t-y)}^{-1}
\eeq quite a bit smaller than the values given by  Eqs
(\ref{Eq;NucBorRates}) and (\ref{Eq;NucPhoProdRates})
respectively.\\

For such axion masses the detection rates in experiments involving
nuclear targets  may be exceedingly difficult. This, of course,
can be traced back mainly to the small Borexino 5.5. MeV axion
flux.


\section{The axion induced processes in atomic targets}

The presence of axion electron interaction of the form of Eq, \ref{Eq:spinae}
gives rise to some interesting processes involving atomic targets. The most interesting are:\\
i) The  process:
\beq
a+A(Z)\rightarrow A'(Z)+\gamma
\label{Eq:Processag}
\eeq
with $ A'(Z) $ the ground state or excited state of the neutral atom. This process is historically
 known as inverse Compton scattering or axion induced Compton scattering.\\
ii) The process
\beq
a+A(Z_e)\rightarrow A^{*}(Z_e-1)+e^-
\label{Eq:Processaef}
\eeq
where $Z_e$ is the number of electrons in the atom. In other words the final atom contains an electron hole.
This process is historically known as axio-electric effect or axion induced atomic excitation.

\subsection{The axion electron coupling}

Before proceeding with the evaluation of the   cross section for
axion induced atomic process, we will consider  the axion electron
coupling, see section \ref{sec:pmodel} which can be written as
\beq {\cal L}=G_e\sbf\cdot {\bf q},\,G_e=\frac{g_{ae}}{2 f_a},\,
\eeq where $g_e$ is a dimensionless coupling constant, which can
be computed in axion models, e.g. \cite{DineFisc83}. It argued in
the context of the DFSZ model
\cite{Srednicki85,CHVV16,RingSaika15} that It can be given by a
relation of the form \beq g_{ae}=\frac{1}{3}\left
(1-\frac{\tan^2{\beta}}{1+\tan^2{\beta}} \right
),\,\tan{\beta}=\frac{\upsilon_2}{\upsilon_1}, \eeq where
$\upsilon_2$ and $\upsilon_1$ are the vacuum expectation values of
the two doublets of the DFSZ  model. These are not known.
 In such scenario the maximum value of $g_{ae}$ can be 1/3, but small values
 corresponding to  the large values $ \tan^2{\beta}$ are perhaps favored,
 as had been the case with supersymmetry. Be that as it may, we have decided to be conservative and adopt
a relatively large value of $\tan{\beta}$, namely
$\tan^2{\beta}=12.9$. This yields the  value
 \beq
 g_{ae}=0.024
 \eeq
 In any case, it will be treated as a parameter to be fixed by experiment. To make this clear,
 in all expressions it is going to be used, there will be a relevant accompanying factor, some
  power of $ \left (\frac{g_{ae}}{0.024}\right )$

  On the
other hand $f_a$ is the axion decay constant, which is related to
the axion mass via Eq. (\ref{Eq:mafaNew}).
This relation is close to that    between $m_a$ and $f_a$ used in the analysis of the
CUORE \cite{CUOREColl13} and Borexino \cite{Borexino12} experiments.
Thus for $f_a=10^7$ GeV we get $m_a=0.6$ eV and,  selecting the value $g_{ae}=0.024$, we find
$$G_e=1.2 \times 10^{-12}\mbox{MeV}^{-1}\left (\frac{g_{ae}}{0.024}\right )\frac{m_a}{0.6\mbox{eV} }=1.0 \times 10^{-11}\mbox{MeV}^{-1}\left (\frac{g_{ae}}{0.024}\right )\frac{m_a}{5\mbox{eV} }$$
Thus the scale of the cross section is
\beq
(\sigma_0)_e=G_e^2=1.0 \times 10^{-22}\mbox{MeV}^{-2}\left (\frac{g_{ae}}{0.024}\right )^2\left (\frac{m_a}{5\mbox{ eV}}\right )^2=3.8 \times  10^{-44}\mbox{cm}^2\left (\frac{m_a}{5\mbox{ eV}}\right )^2\left (\frac{g_{ae}}{0.024}\right )^2
\label{Eq:LimAve}
\eeq
We should mention at this point that all cross sections calculated below  will remain the same,  when   $g_{ae}$
and $m_a$ are suitably scaled. For example  if $g_{ae}$ happens to be a thousand times smaller,
 $g_{ae}=2.4\times 10^{-5}$, provided, of course, that
 the axion mass is three orders of magnitude larger, i.e. $m_a=5$ keV.

The 14.4 keV axion source appears to be ideal for atomic physics
experiments. The 14.4 keV  source  is monochromatic, produced in
the sun via the $^{57}$Fe* \cite{SolFlux20}, with a flux given by
Eq. (\ref{Eq:Fe57SF}).

\subsection{A brief summary of limits on the parameters $m_a$ and $f_a$ extracted from experiments}

 In all the calculations of the type we have performed  a crucial parameter is the axion mass $m_a$,
 which is not known. We have been using a value of $m_a=5$ eV, but both lower and  higher values are not excluded.
  Limits much smaller than 5 eV,  of the order of meV, typically come form
  astrophysics \cite{AxionCompton97} (cooling of neutron stars etc via axion emission) or
   supernova (SN) explosions. Recently  bounds come from  (SN) data \cite{AxionBrem19} which
   vary from 10 to 50 meV in the context of KVSZ model. Such light axions are, of course,
   not relevant for nuclear and atomic processes considered in this
   work.

  Higher bounds have also been given:\\
  a) The CUORE \cite{CUOREColl13} collaboration using  the 14.4 keV solar axions  and
  measuring  the axio-electric effect in the TeO$_2$ bolometers,
  provides two limits on $m_a$:
    \beq
    m_a \le 19.2 \mbox{ eV and } m_a \le 250 \mbox{ eV both at } 95\% \mbox{ C.L.}
    \label{Eq:CUOREbounds}
    \eeq
    Strictly speaking what  the experiment yielded  are limits on $f_a$ by using the DFSZ and KSVZ model for $g_{Ae}$, i.e.
     values of $f_a\ge3.125\times 10^5$ GeV and $\ge2.4\times 10^4$ GeV respectively. Then they extracted  the above values of $m_a$  as implied by Eq. (\ref{Eq:mafaNew}).\\
  b) Axions of energy 14.4 keV were employed using the axion induced excitation of $^{57}\mbox{Fe}$, the
  inverse of the axion production in the sun. That is the   process $a+^{57}\mbox{Fe}\rightarrow  ^{57}\mbox{Fe}^* \rightarrow ^{57}\mbox{Fe}+\gamma  \mbox{(14.4 keV)}$ with a
  sectioned Si(Li) detector arranged in a low-background facility \cite{DerNurSemUnz11}.
  This approach  succeeded in setting a new limit on the axion couplings to nucleons,
  $ |-1.19g^0_{aN} + g^3_{aN}|\le 3.0\times10^{-6}$. Within the hadronic-axion model,
  the respective constraint on the axion mass is $ m _a \le 145$ eV (at a $95\%$ C.L.).
   An analysis of the CUORE sensitivity to 14.4 keV solar axion  detected by inverse coherent Bragg-Primakoff
   conversion in single-crystal TeO$_2$ has been performed. The axion coupling of $g_{a\gamma\gamma}g^{eff}_{aN}$
   to be studied by CUORE was calculated (evaluated ) to be $<1.105\times 10^{-16}$/GeV. In this analysis   an
    axion mass of less than  500 eV has  been used  \cite{LiAvigWang16}.

  In another search for solar axions \cite{AxionBremCom11} produced by Compton ($\gamma+e^-\rightarrow e^-+a$ )
  and bremsstrahlung- like ( $e^-+Z \rightarrow e^- +Z+a$) processes has been performed. In this case  an
  upper limit on the axion-electron coupling and on the axion mass, i.e. on $g_{Ae} \times  m_a \le 3.1 \times 10^{-7}$
  eV at (90$\%$C.L.), has been obtained. The limits on axion mass are $m_a\le 105$ eV and
   $m_a\le 1.3$ keV at the  (90$\%$ C.L.) for DFSZ and KSVZ axion
  models, respectively.

   The existence of   high upper bounds on the axion mass are extremely encouraging
   to us in connection with this work, since they allow the possibility detectable rates
   involving axions in the 5 eV range, i.e. $f_a$ around $10^6$ GeV.

  We stress that  there exist two components in our calculation.
  First the size of the oncoming axion flux, coming from the production mechanism,  which can now be written as
  \beq
  \Phi(E_a)=0.7 \times 10^9 \left (\frac{m_a}{0.6 \mbox{ eV}}\right )^2\mbox{cm}^{-2}\mbox{s}^{-1} =2.2\times 10^{16} \left (\frac{m_a}{0.6 \mbox{ eV}}\right )^2\mbox{cm}^{-2}\mbox{y}^{-1}\mbox{ (14.4 keV solar axion)}
  \label{Eq:Fe57SF2}
  \eeq
  This, for convenience, can be re-scaled
  \beq
  \Phi(E_a)=1.5\times 10^{18} \left (\frac{m_a}{5 \mbox{ eV}}\right )^2\mbox{cm}^{-2}\mbox{y}^{-1}\mbox{ (solar for 14.4 keV axion, rescaled)}.
  \label{Eq:amendedflux}
  \eeq
  The value for $m_a=5$ eV  corresponds to $f_a=1.2 \times 10^6$
  GeV.

   Second,  the calculated cross section, which is also proportional to $m^2_a$. These combined
    for $N_A=4 \times 10^{27}$ particles in the target (1ton of NaI) yield a rate scaled by
 \beq
 R_0=1.5 \times 10^{18} \times 4 \times 10^{27}\times  3.8\times 10^{-44} \left( \frac{m_a}{5 \mbox{eV}}\right )^4=228 \mbox{ events per t-y}\left( \frac{m_a}{5 \mbox{eV}}\right )^4
 \label{Eq:AxionScRate}
 \eeq

 It seems that in the atomic experiments one should use the value of $\sigma_0$ given
 by Eq. (\ref{Eq:LimAve}) and the flux given by
(\ref{Eq:amendedflux}).

 \section{Some results on  Axion electron induced transitions in the case N$a$I  target}

 In our calculations we  will  consider the following processes:

 \subsection{Axion photoproduction on electrons}

 This process involves a special mechanism  leading to the reaction  given by Eq. (\ref{Eq:Processag}). In other words  in this case the process can proceed in a manner analogous to Fig. \ref{fig:Nga}(b), with the obvious replacements $N\rightarrow e$, $g^3_{aN}\rightarrow g_{ae}$. It involves the spin dependent axion electron interaction  as  given by Eq. (\ref{Eq:spinae}) and the magnetic moment of the electron in the photon vertex. The  name came from nuclear physics, the old analogous pion photoproduction process. In the atomic case it is a special case of the axion induced Compton scattering. It involves the gs or low lying atomic excitations. Other possible mechanisms leading to reaction  given by Eq. (\ref{Eq:Processag}),   involving, e.g., the
  inverse Primakoff axion scattering \cite{LiAvigWang16,AbeMamNag21}, will not be considered.

  Let us estimate the expected cross sections, by  comparing it  to the  hadronic process. The new feature is
   the Bohr magneton for the electron (as opposed to the nucleon one used before) and the axion
   energy dependence. Thus for the 14.4 keV  axion we get
 \beq
 \sigma_0=\frac{m^2_n}{m^2_e} \left (\frac{14.4\mbox{keV}}{5.5\mbox{MeV}}\right )^2 \times 6.0\times10 ^{-50} =1.5\times 10^{-48}\mbox{cm}^{2}
 \label{Eq:sigma0Cox}
 \eeq
 see Eq. ( \ref{Eq:sgma0RadCap}). This, however, has been estimated with $f_a=10^7$ GeV. Thus,  in view of Eq. (\ref{Eq:mafaNew}),  it can be written as
 \beq
  \sigma_0=1.5\times 10^{-48}\mbox{cm}^{2}\left (\frac{m_a}{0.6 \mbox{eV}}\right )^2=1.1 \times 10^{-46}\mbox{cm}^{2}\left (\frac{m_a}{5 \mbox{eV}}\right )^2
  \label{Eq:sigma0Cox2}
 \eeq
 The dimensionless couplings will be included separately.\\
 For the 5.5 MeV axion we have
 $$\sigma_0=\frac{m^2_n}{m^2_e} \times 6.0\times 10^{-50}=2.1 \times 10^{-43}\mbox{cm}^{2}$$
 The second process, of course, is not relevant, since,  as we have seen, with the Borexino flux it is    undetectable.
 There is, of course, a coherence factor, that goes with the electron axion - photoproduction, which  is now proportional to $Z^2$.
  Thus  for 14.4 keV axion  proceeding as in the nuclear axion photoproduction we find:
 \beq
 \sigma(E_a)=\sigma_0 G^2_{ae\gamma}\left(\mbox{ME}_{el}^2+\frac{1}{3}\frac{1}{2 J_i+1}\sum_{f}\langle J_f||\sbf||J_i\rangle^2\left(1-\frac{Ex}{E_a}\right )^3\right ),
 \eeq
 with $\sigma_0$ given by Eq. (\ref{Eq:sigma0Cox2}) and  the effective axion-electron-photon coupling,
  analogous to those of Eq. (\ref{Eq:GPhoto}), is
 \beq
 G_{ae\gamma}=g_s g_{ae}=2\times 0.024 =0.048
 \eeq
 \beq
 \mbox{ME}_{el}^2=\left \{\frac{2}{3} Z^2+\frac{1}{3}\frac{1}{2 J_i+1}\langle J_i||\sbf||J_i\rangle^2  \right \}.
 \eeq
 Where in the atomic system $E_x<< E_a$.\\
 Compare this formula with the analogous expression in nuclear transitions, Eq, (\ref{Eq:elasticNucleus}),
 and notice that in the present case the spin independent operator, see the analogous operator  $\Omega_A$
  in Eq. (\ref{Eq:PhoprodOp}), is just a multiple of the identity,  yielding the $Z^2$ dependence of the
  square of the relevant matrix element.

 The spin reduced ME for single particle transitions are given by:
 \beq
 \left(
 \begin{array}{ccc|c}
    \ell&j&j'&\langle \ell,j'||\sigma||\ell,j\rangle^2\\
    \hline
    0 & \frac{1}{2} & \frac{1}{2} &
    6 \\
    1 & \frac{1}{2} & \frac{3}{2} &
    \frac{16}{3} \\
    1 & \frac{1}{2} & \frac{1}{2} &
    \frac{2}{3} \\
    1 & \frac{3}{2} & \frac{1}{2} &
    \frac{16}{3} \\
    1 & \frac{3}{2} & \frac{3}{2} &
    \frac{20}{3} \\
    2 & \frac{3}{2} & \frac{5}{2} &
    \frac{48}{5} \\
    2 & \frac{3}{2} & \frac{3}{2} &
    \frac{12}{5} \\
    2 & \frac{5}{2} & \frac{3}{2} &
    \frac{48}{5} \\
    2 & \frac{5}{2} & \frac{5}{2} &
    \frac{42}{5} \\
    3 & \frac{5}{2} & \frac{7}{2} &
    \frac{96}{7} \\
    3 & \frac{5}{2} & \frac{5}{2} &
    \frac{30}{7} \\
    3 & \frac{7}{2} & \frac{5}{2} &
    \frac{96}{7} \\
    3 & \frac{7}{2} & \frac{7}{2} &
    \frac{72}{7} \\
 \end{array}
 \right)
 \eeq
Even for Na, the lighter component of the target, we see that the  coherent contribution becomes dominant. \\
 Thus the obtained coherent  rates, obtained with the flux  given by
 (\ref{Eq:amendedflux}), are
\barr
R_{I}&& \approx 2.78 \left (\frac{g_{ae}}{0.024}\right )^2\left (\frac{m_a}{5 \mbox{ eV}}\right )^4\mbox{ events per (t-y)},\,R_{Na} \approx 0.097 \left (\frac{g_{ae}}{0.024}\right )^2\left (\frac{m_a}{5 \mbox{ eV}}\right )^4\mbox{ events per (t-y)}\Rightarrow\nonumber\\  R&&\approx 2.92 \left (\frac{m_a}{5 \mbox{ eV}}\right )^4\left (\frac{g_{ae}}{0.024}\right )^2\mbox{ events per (t-y)}
\label{Eq:PhotProdt}
\earr
In this mechanism the photons are radiated off electrons directly. As far as we know, nobody has previously
 considered this axion photoproduction  mechanism.

The above expression does not depend on the details of the atomic structure of the target, but due to
 coherence only to the atomic number. Also this process leads directly to photons with an energy equal to
 the axion energy $E_a$. Recall, however that the scale of the cross section  is  proportional to  $E^2_a$.
 Recalling  Eq.  (\ref{Eq:AxionScRate}) we get a sort of global expression:
\beq
R(Z)\approx 2.50 \left (\frac{g_{ae}}{0.024}\right )^2\left (\frac{Z}{50}\right )^2\left (\frac{Ea}{14.4 \mbox{keV}}\right )^2 \left( \frac{m_a}{5 \mbox{eV}}\right )^4\mbox{ events per y}
\label{Eq:RateR(Z)}
\eeq
for $N_A=4 \times 10^{27}$

\subsection{ Ejection of an electron from an interior orbit}

This process is described by Eq,  (\ref{Eq:Processaef}), but with 14.4 keV axion source the electron hole can be in
 a deep bound orbit and as result the  outgoing electron energy is  $E_a+\epsilon_b$. Such an electron can be observed directly.
 In addition one  may observe the X-rays and Auger electrons  following the de-excitation of the final atom.

 We will now proceed with estimates of the total cross section. The essential mathematical steps are given
 in the Appendix, section \ref{sec:Appendix}.\\
 For bound  electrons the cross section depends on two parameters: a) the form factor like  dependence
 involving the bound state wave function in momentum space  $F_{n\ell}=a^3\Phi^2_{n \ell}(a,k)$, with $\Phi_{n \ell}(a,k)$
 the electron wave function in momentum space. Here ${\bf k}=\bf{q-p}_e$ is the momentum transfer  with  ${\bf q}$ the
  axion momentum and ${\bf p}_e$ the momentum of the outgoing electron and b) The phase space factor $K_n$, which has
   been evaluated with the same binding for all states in a given $n$. Hydrogenic bound state wave functions we
    employed with $a=\frac{\alpha Z m_e}{n}$. The function  $F_{n\ell}$ for convenience has been chosen to be
     dimensionless  with a compensating factor $a^{-3}$ in the phase space part to make it also  dimensionless.

  \subsection{The cross section for ionization of the Iodine Component}

 Bound states up to $n=4$ were considered, with the last one being partially filled.
 The form factor in the case of the 14.4 keV axions is dominated by the outgoing electron
  momentum $p_e=\sqrt{2 m_e (E_a+\epsilon_{b})}$.
 The corresponding electron energies are:
 $$T=\{4.84, 10.15, 12.01\}\mbox{ keV for }n=2,3,4\mbox{ resectvely}$$
 In the case of the 5.5 MeV axions the situation is a bit complicated and we made the simplifying
 assumption $k=\sqrt{q^2+p_e^2}$. In this case the binding energy becomes irrelevant and the electron
 momentum $p_e=\sqrt{E_a(E_a+2m_e)}$.

 We should mention that we summed over all states in a given full shell by introducing the $(4\ell+2)$ factor.
  The form factor $F_{n\ell}$ and the  kinematical factor $K_n $ are given as follows:
 \beq
 F_{n\ell}=\left(
 \begin{array}{cccccc}
 n&K(n)&F_S(n)&F_P(n)&F_D(n)&F_F(n)\\
 \hline
 2&0.536&1.69994 & 14.1273 & - & - \\
 3&2.619&0.0166029 & 0.517959 &
 0.854939 & - \\
4&6.7547&0.00127243 & 0.0770985 &
 0.135348 & 0.0335818 \\
\end{array}
\right)
 \eeq
The  $S_{n\ell}$ factors for the cross section are given by
\beq
S_{n\ell}=\left(
\begin{array}{ccccc}
    n&S(S(n))&S(P(n))&S(D(n))&S(F(n))\\
    \hline
    2&  0.911049 & 7.57128 & - & - \\
    3&  0.0434898 & 1.35675 & 2.23944 & - \\
    4&  0.00859345 & 0.52069 & 0.914085 & 0.226797 \\
    \end{array}
\right)
    \label{Eq:SnlOr}
\eeq
 The corresponding form factor in the case of 5.5 MeV energy is given by:
 \beq
 F_{n\ell}=\left(
 \begin{array}{ccccc}
         n&F_S(n)&F_P(n)&F_D(n)&F_F(n)\\
\hline
 1& 3.5376 \times 10^{-12}& 0 & 0 & 0 \\
 2& 5.5345\times 10^{-14} &2.3850 \times 10^{-17} & 0 & 0 \\
 3& 4.8600 \times 10^{-15}&2.4813 \times 10^{-18} &2.1373 \times 10^{-22}& 0 \\
 4& 4.6504\times 10^{-16} &4.6580\times 10^{-19} &5.4167 \times 10^{-23} &1.6663\times 10^{-27} \\
 \end{array}
 \right)
 \eeq
 The total $S$ factor for the cross section  is given by
 \beq
 S_{n\ell}=\left(
 \begin{array}{ccccc}
    n&S(S(n))&S(P(n))&S(D(n))&S(F(n))\\
 \hline
 1& 7.6183\times 10^{-8} &  0 & 0 & 0 \\
 2& 9.53514 \times 10^{-9} &    4.1075 \times 10^{-12} & 0 & 0 \\
 3& 2.8259 \times 10^{-9} &1.4428 \times 10^{-12}   &   1.2428 \times 10^{-16}  & 0 \\
 4& 1.1923 \times 10^{-9} &6.4200\times 10^{-13} &7.4657 \times 10^{-17} &2.2966\times 10^{-21} \\
 \end{array}
 \right)
 \eeq
 We see  that the largest $S_{n \ell}$-factor is   $7.6\times 10^{-8}$ coming for the ejection of an
  $0s$ electron. The  form factor for this large momentum transfer leads to a great suppression of the cross section,
  in spite of the fact that the kinematical factor is favored, $K_n=0.022,\,0.17,\,0.58,\,1,37\times 10^{6}$, for $1,2,3,4$
   respectively.
 In addition  the 5.5 MeV axions are not favored in atomic physics experiments due to the very small flux involved.

 In the case of the 14.4 keV axions the cross section is
 \beq
\sigma_{n,\ell}=\sigma_0 \frac{1}{2 \pi}S_{n,\ell}
\label{Eq.StoSigma}
 \eeq
 (see Eq. ( \ref{Eq:sigma0S}),
 with $\sigma_0$  given by Eq. (\ref{Eq:LimAve}). Thus we get
 \beq
  \sigma_{n,\ell}=\{ 0.549, 4.580, 0.0279, 0.8200, 1.353, 0.005,
  0.314, 0.553, 0.135\} \times 10^{-44}\mbox{cm}^2\left (\frac{g_{ae}}{0.024}\right )^2\left( \frac{m_a}{5 \mbox{eV}}\right )^2
 \label{Eq:sigmanl}
 \eeq
in the order of the orbitals appearing in Eq.  (\ref{Eq:SnlOr}).

 \subsection{The cross section for ionization of  the Na Component}

\begin{figure}[htbp]
    \begin{center}
         \subfloat[]
        {
            \includegraphics[width=0.4\textwidth]{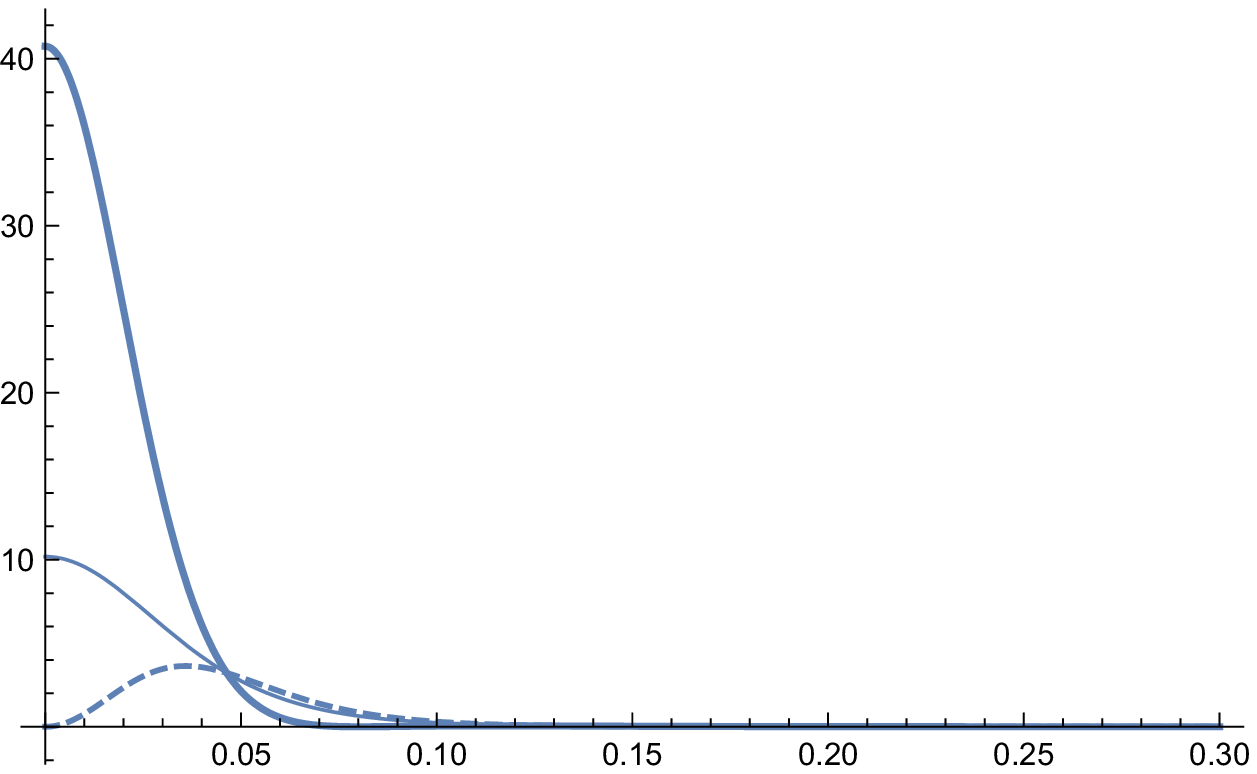}
        }
         \subfloat[]
        {
            \includegraphics[width=0.4\textwidth]{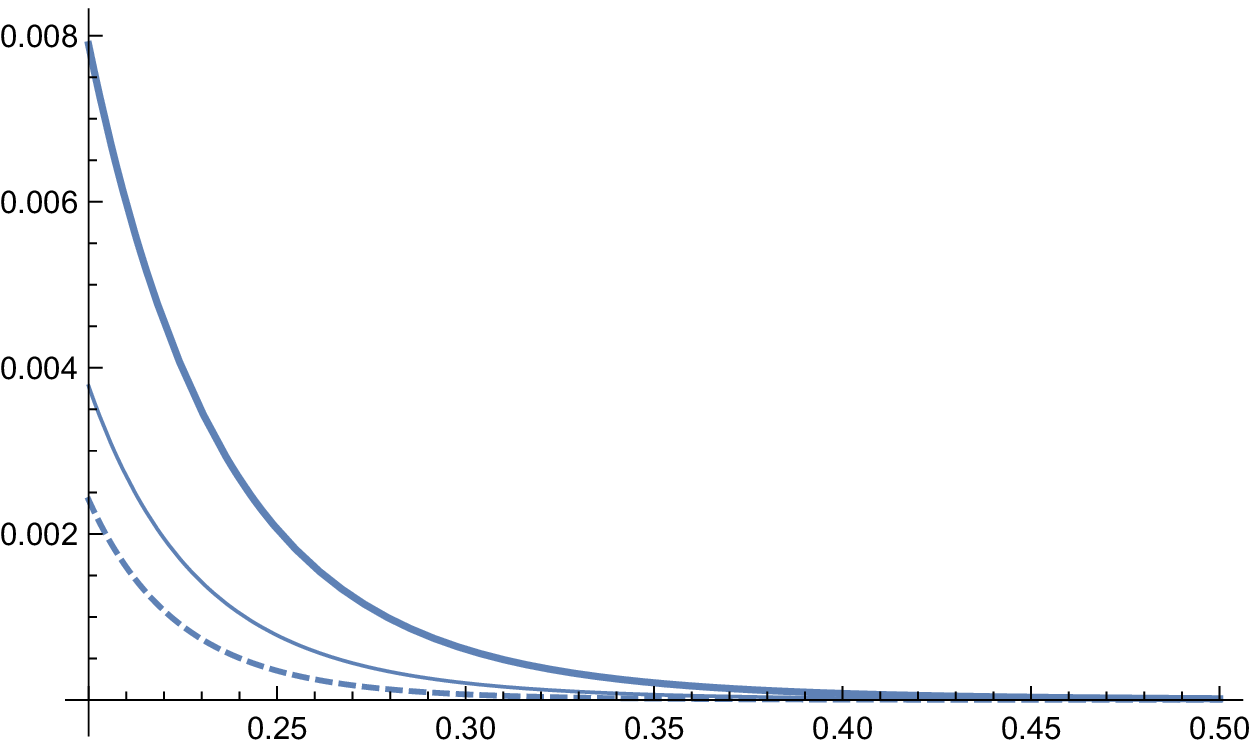}
        }
        \\
        {\hspace{-2.0cm} {$X=\frac{P_e}{m_{e}}\longrightarrow$}}\\
        \caption{(a) We exhibit the dependence of the form factors encountered in the ionization of Na as a function of the momentum transfer, here momentum  of the emitted electron $p_e$ in units of the electron mass. (b) The same figure restricted to higher momentum transfers. The solid line corresponds to $n=1$, the thick solid line to $n=2,\ell=0$   and the dashed line to  $n=2,\ell=1$.}
        \label{fig:PlotFF}
    \end{center}
 \end{figure}

 We do not expect a large cross section in this case because the energy of the
 outgoing electron is larger  due to the smaller binding energy involved. Thus
 one expects a lager damping on the form factors  of the available $n=1$ end $n=2$
 orbitals.
 In this case we have $\epsilon_n=(-1.646,-0.420)$keV respectively for $n=1,2$ .
 On the other hand the outgoing electron  energies are:
 $$T=\{12.75, 13.99\}\mbox{ keV for }n=1,2\mbox{ resectvely}$$
 The suppression of the form factors at high  momentum  transfer, here essentially the momentum of the outgoing electron,
     can be seen in  Fig.  \ref{fig:PlotFF} as a function of $X$, the momentum transfer in units of the electron mass.
     The actual electron momenta involved here are    $X=(0.2234, 0.2340)$ for electrons ejected from the $n=1$ and $n=2$
     respectively.

  In view of this,  it is not surprising that, in spite of the fact the kinematical function $K_n$ is
  quite large,  the  $S_{n\ell}$ factors are small:
 \beq
 S_{n\ell}=\left(
 \begin{array}{cccc}
    n&K(n)&S(S(n))&S(P(n))\\
    \hline
    1&12.2& 0.0424184 & - \\
    2&101.9&    0.00493614 & 0.000617041 \\
    \end{array}
    \right)
 \label{Eq:SnlOrNa}
 \eeq
 Thus using Eq. (\ref{Eq.StoSigma}) we find
 $$
 \sigma_{n\ell}= \{2.567, 0.298, 0.036\}\times 10^{-46}\mbox{cm}^2\left (\frac{g_{ae}}{0.024}\right )^2\left( \frac{m_a}{5 \mbox{eV}}\right )^2
 $$
The total cross section is
\beq
\sigma=\sum_{n\ell}\sigma_{n\ell}=8.36 \times 10^{-44}\mbox{cm}^2\left (\frac{g_{ae}}{0.024}\right )^2\left( \frac{m_a}{5 \mbox{eV}}\right )^2
\eeq
This cross section scaled to $m_a=5$keV becomes $8.36 \times 10^{-38}\mbox{cm}^2$ which between the
values $10^{-36}\mbox{cm}^2$ (DFSZ) and  $10^{-41}\mbox{cm}^2$ (KSVZ) reported by CUORE \cite{CUOREColl13}.

 \subsection{Some results regarding the rates for atomic ionization}
 \label{sec:eresults}

  Using now the 14.4 keV  flux given by Eq.
  (\ref{Eq:amendedflux})) and $N_A=4 \times 10^{27}$ as above we find:

  i) For the iodine component:
 \beq
 R_{n,\ell}=
\{ 0.337, 2.810, 0.016, 0.503, 0.830, 0.003,  0.194, 0.340, 0.082\}\times 10^2
 \mbox{per t-y}\left( \frac{m_a}{5 \mbox{eV}}\right )^4\left (\frac{g_{ae}}{0.024}\right )^2
 \label{Eq:AtIonRateI}
 \eeq
 in the same order as in Eq,  (\ref{Eq:sigmanl}).

 ii) For the Na component
 \beq
 R_{n,\ell}=\{1.575, 0.182, 0.022\}\mbox{per t-y}\left( \frac{m_a}{5 \mbox{eV}}\right )^4\left (\frac{g_{ae}}{0.024}\right )^2
  \label{Eq:AtIonRateNa}
 \eeq

 The obtained rates depend on the atomic wave function and in particular the form
 factor of the orbit involving the ejected electron. We have seen that a great suppression
 of the form occur, if the axion energy is high, e,g,  5 MeV. The same thing happens
 in the case of a 14,4 keV axion for a small $Z$ atom, since again, due to the small binding,
 the form factor is suppressed due large momentum transfer, see Fig.  \ref{fig:PlotFF}. Thus,
 for the Iodine atom, as well as others with nearby $Z$ and the 14.4 keV energy, we have a
  favorable situation. One can write for the dominant contributions:
  \beq
 R_{n,\ell}=\{0.337, 2.810, 0.503, 0.829, 0.340\}\times 10^2 \mbox{per t-y}\left( \frac{m_a}{5 \mbox{eV}}\right )^4\left (\frac{g_{ae}}{0.024}\right )^2
 \label{Eq:AtIonRateIlim}
 \eeq
 in the order $2S,\,2P,\,3P,\, 3D,\mbox{ and }4D$ for  the orbits.\\
 We notice the dramatic dependence of the event rate on the axion mass. Deceasing the axion mass by a
 factor of 5 will lead to a reduction of the rate by 3 orders of magnitude and make it un-observable,
  an increase by the same factor makes one wonder how come such an axion has not already  been seen.

\section{Concluding remarks}

In the present work we studied various axion induced nuclear and
atomic processes, assuming  an axion mass of 5 eV.

 In the case of
the nuclear processes we considered the mono-energetic axion
source of 5.5 MeV. The obtained nuclear cross sections were found
to be reasonable, but the obtained rates were very small mainly
due to the small flux on the earth of such axions,  obtained with
the Borexino detector \cite{Borexino12}. So, such processes are
doomed to be non detectable for axion masses in the few eV range.
Had the mass of the axion been in the keV region, such processes
would become detectable with rates as given by Fig. \ref{rvma}
and, in particular, by Eqs ~(\ref{Eq;NucBorRates}) and
(\ref{Eq;NucPhoProdRates}). We have seen, however, that such
masses are excluded by astrophysical data.
 For lower axion masses in the subkeV region,  e.g.  the value of  $m_a=0.25$keV,
  a model dependent limit extracted by the CUORE experiment \cite{CUOREColl13},
  the obtained total excitation  and photoproduction rates in the case of  NaI
  target become  $R\approx 6\times 10^{-3}$ and $R_{coh}\approx 10^{-5} \mbox{(t-y)}^{-1}$
   respectively, exceedingly difficult to detect.

In the case of atomic experiments, with inner  electron binding
energies
 in the keV region, we found  appropriate and we  used the  14.4 keV $^{57}$Fe axion  source.
 In this case we find sizable rates, e.g.:\\
i) Axion induced photoproduction. We get the rate given by Eq. (\ref{Eq:RateR(Z)})
 for directly producing  X-rays with energy equal to the axion energy $E_a$.
  Smaller rates are expected from the excited states with about the same energy.
  This is independent of the details of the structure of the atom. For $m_a=5$ eV axions ($f_a=1.2 \times 10^{6}$ GeV)
   this leads to  a rate $ R\approx 2.50\mbox{ events per (t-y)}$ in the case of NaI  target, assuming a coupling $g_{ae}=0.024$.\\
ii) Electron ejection from the atom. In this case we find sizable
rates for ejecting an electron from the relevant  orbits of the
Iodine, given by Eq.  (\ref{Eq:AtIonRateIlim}), the maximum being
280  events per t-y for $m_a=5$ eV ( $f_a=1.2 \times 10^{6}$ GeV),
again for $g_{ae}=0.024$, coming from the $n=2,\ell=1$ orbit, and
slightly smaller rates from some of the other orbits. Much smaller
rates  are expected for the Na component.

Experiments like those discussed above  have already been done by CUORE-Te  \cite{CUOREColl13} and others \cite{LiAvigWang16}, \cite{DerNurSemUnz11}.
As given in Eq.(\ref{Eq:CUOREbounds}), the 14.4 keV axion with the rate larger than $R= 2\times 10^5(t- y)^{-1} (m_a<19$ eV DFSZ) has been excluded \cite{CUOREColl13}. Then,  axions in the  region of  $R=(10^{2}-10^{4})(t-y)^{-1}$, i.e. $m_a=5-10$ eV, $g_{ae}=0.024$,  are very interesting and  their detection  in the case of atomic excitations is quite realistic, by improving detector sensitivities by a few orders of magnitude.


 If that is the
case, we expect a very interesting axion signature, i.e. a signal
consisting of the detection of the primarily produced electrons
from each orbit  as well as the detection of the X-rays or Auger
electrons produced, when the created hole is filled by the
de-excitation of the atom. The relevant rates depend, of course,
on the assumed value of the elementary coupling $g_{ae}$, which is
not really known.

 \section{Appendix: The formalism of  the axion electron cross section from bound orbits}

\label{sec:Appendix}
The axion electron scattering  cross section for relativistic axions is given by
\beq
d \sigma =\sigma_0\int {\cal M}(k^2)^2\frac{1}{2 E_a}\frac{d{\bf p}_ e^3}{(2 \pi)^3}\frac{d{\bf p}_A ^3}{(2 \pi)^3}(2 \pi)\delta(E_a+\epsilon_ b-T)(2 \pi)^3\delta ({\bf p}_a-{\bf p}_A-{\bf p}_e)
\eeq
where ${\cal M}(k^2)$ is the invariant amplitude, which depends on the momentum transfer, ${\bf p}_ e$ and $T$ are the momentum and the energy of the outgoing electron and  ${\bf p}_ A$ of the atom, $\epsilon_ b $ is the binding energy of the electron in the target. $E_a$ is the energy of the axion, normally being  introduced as a normalization of the scalar field. We have neglected the energy of the outgoing atom. Thus integrating over ${\bf p}_ A$ with the use of the momentum conserving $\delta$ function we obtain
\beq
 \sigma =\sigma_0\frac{1}{2}\frac{1}{(2 \pi)^2} 4 \pi S=\frac{1}{2 \pi}S,\,S=\frac{1}{4 \pi}\int {\cal M}(k^2)^2\frac{1}{ E_a}d{\bf p}_ e^3\delta(E_a+\epsilon_ b-T)
 \label{Eq:sigma0S}
\eeq
 Let us begin with the discussion of ${\cal M}(k^2)^2$. The relevant matrix element (ME) involving  the initial bound electron and the outgoing plane wave electron is given by:
 \beq
 \mbox{ME}=\int \psi_{n\ell m}({\bf r})e^{i{\bf q\cdot {\bf r}}}\langle \frac{1}{2}m_s|\sbf\cdot{\bf q}|\frac{1}{2}m_s'\rangle  \frac{1}{(2\pi)^{3/2}}e^{-i{\bf p_e\cdot {\bf r}}} =\phi_{n\ell m}({\bf k})\langle \frac{1}{2}m_s|\sbf\cdot{\bf q}|\frac{1}{2}m_s'\rangle ,\,{\bf k}={\bf q}-{\bf p}_e
 \eeq
 Where $\phi_{n\ell m}({\bf k})$ is the initial electron bound wave function in momentum space. The last term involves the axion electron interaction.We thus find that transition probability averaging over the initial m-substates and summing over the final ones becomes:
 \beq
 \frac{1}{( 2 j+1)}\sum_{m,m_{\ell},m_s.m_s'}\langle \ell m_{\ell},\frac{1}{2}m_s|j m\rangle^2 \mbox{ME}^2=|\phi_{n\ell m}({\bf k})|^2\langle \frac{1}{2}\sum_{m_s,m_s'}\frac{1}{2}m_s|\sbf\cdot{\bf q}|\frac{1}{2}m_s'\rangle^2
 \eeq
 The last result was obtained using the properties of the Clebcsh-Gordan coefficients.The last matrix element can easily be evaluated yielding $q^2 \approx E_a^2$   We thus find
 \beq
 S=\frac{1}{4 \pi}\int |\phi_{n\ell m}({\bf k})|^2 E_a d{\bf p}_ e^3\delta(E_a+\epsilon_ b-T)
 \eeq
 We will specialize it in the case of 14.4 keV axion. In this case the outgoing electron is non relativistic $T=p_e^2/(2 m_e)$, with the use of $\delta$ function we  get $p_e=\sqrt{2m_e(E_a+\epsilon_b)}$, which is much larger than the axion momentum. So $k\approx p_e$.  So after the integration over the angles we get:
 \beq
 S=\left (\phi_{n\ell m}(\sqrt{2m_e(E_a+\epsilon_b)})\right )^2 E_a m_e\sqrt{2m_e(E_a+\epsilon_b)}
 \label{Eq:FnlFF}
 \eeq
 At this point we recall that the bound state wave function can be written as $\phi_{n \ell}(a,k)$, where $k$ is the momentum and $a=\frac{\alpha Z m_e}{n}$. We prefer to write it in dimensionless form and write is as
 \beq
 F_{n \ell}(a,k)=a^3 \phi^2_{n \ell}(a,k)\Rightarrow S= F_{n \ell} K_n,\, K_n=n^3\frac{1}{(\alpha Z)^3}\frac{E_a}{m_e}\sqrt{2 \left (\frac{E_a+\epsilon_b(n)}{m_e}\right )}
 \eeq
 A compensating $1/a^3$ factor has been introduced in the kinematical portion of Eq, (\ref{Eq:FnlFF}). This way both  $F_{n \ell}$ and $ K_n$ are dimensionless.

 In the case of the 5.5 MeV axion we proceed in  an analogous way. Now the binding becomes irrelevant and the outgoing electron caries all the energy, Its momentum however is $\sqrt{E_a(E_a+2m_e)}$, approximately the same with the axion momentum. So the angular integral is quite complicated and we made the simplifying assumption $k=\sqrt{q^2+p_e^2}$
 Now
 \beq
 K_n=n^3\frac{1}{(\alpha Z)^3}\left (\frac{E_a}{m_e}\right )^3
 \eeq
 As expected $ K_n$ becomes large, but the form factor  $F_{n \ell}(a,k)$ at this high momentum transfer is tiny, so that the cross section becomes  negligibly small.

 The relevant for factors employed are given below:
 $$
 F_{1s}=\frac{128 a^8 (a-k)^2
    (a+k)^2}{\pi
    \left(a^2+k^2\right)^6},\,F_{0p}=\frac{512 a^{10} k^2}{3 \pi
    \left(a^2+k^2\right)^6},\,F_{2s}=\frac{32 a^8 \left(3 a^4-10 a^2
    k^2+3 k^4\right)^2}{\pi
    \left(a^2+k^2\right)^8}
 $$
 $$
 F_{1p}=\frac{1024 a^{10} k^2 (a-k)^2
    (a+k)^2}{\pi
    \left(a^2+k^2\right)^8},\,F_{0d}=\frac{4096 a^{12} k^4}{5 \pi
    \left(a^2+k^2\right)^8}
,\,F_{3s}=\frac{512 a^8 \left(a^6-7 a^4
    k^2+7 a^2
    k^4-k^6\right)^2}{\pi
    \left(a^2+k^2\right)^{10}}
 $$
 $$
 F_{2p}=\frac{2048 a^{10} k^2 \left(5
    a^4-14 a^2 k^2+5
    k^4\right)^2}{15 \pi
    \left(a^2+k^2\right)^{10}},\, F_{1d}=\frac{32768 a^{12} k^4
    \left(a^2-k^2\right)^2}{5 \pi
    \left(a^2+k^2\right)^{10}},\,F_{0g}=\frac{131072 a^{14} k^6}{35 \pi
    \left(a^2+k^2\right)^{10}}
 $$

{\bf Acknowledgments:} J. D. V. would like to thank Georg Raffelt for his help in getting the
axion mass dependence of the 14.4 keV axion solar flux.

\end{document}